\documentclass{article}

\usepackage{xcolor}

\newcommand{\blu}{\color{blue}}

\def\giorno{1/10/2021}

\def\gcite#1{{\blu \cite{#1}}}

\def\symmref{AVL,CGbook,KrV,Olver1,Olver2,Stephani,Win}
\def\sderef{Arnold,Evans,Fre,Ikeda,Kampen,Oksendal,Stroock}
\def\dsref{GH,Gle,IA,Ver}
\def\stochsymmref{GRQ1,GRQ2,Unal,Koz1,Koz2,Koz3,Koz12,GS17,GS17E,GGPR,GGPRE,GL1,GL2,Koz18a,Koz18b,KozB,GLS,GW18}

\def\a{\alpha}
\def\b{\beta}
\def\ga{\gamma}
\def\de{\delta}   
\def\eps{\varepsilon}
\def\vphi{\varphi}
\def\la{\lambda}

\def\s{\sigma}

\def\om{\omega}

\def\vphi{\varphi}

\def\G{{\mathcal G}}

\def\L{\mathcal{L}}

\def\R{{\bf R}}

\def\T{{\rm T}}

\def\Ga{\Gamma}
\def\De{\Delta}

\def\pa{\partial}

\def\xb{{\bf x}}

\def\wb{{\bf w}}

\def\o+{\oplus}
\def\xd{{\dot x}}
\def\yd{{\dot y}}
\def\ss{\subset}
\def\sse{\subseteq}

\def\<{\langle}
\def\>{\rangle}

\def\({\left(}
\def\){\right)}
\def\[{\left[}
\def\]{\right]}
\def\=#1{\bar #1}
\def\~#1{\widetilde #1}
\def\wt#1{\widetilde #1}
\def\.#1{\dot #1}
\def\^#1{\widehat #1}

\def\"#1{\ddot #1}

\newcommand{\beql}[1]{\begin{equation}\label{#1}}
    \newcommand{\beq}{\begin{equation}}
    \newcommand{\eeq}{\end{equation}}   
    \def\eqref#1{(\ref{#1})}
    
    \def\EOR{{ } \hfill $\odot$ \medskip}
    \def\EOE{{ } \hfill $\diamondsuit$ \medskip}
    \def\EOP{{ } \hfill $\diamondsuit$ \medskip}

    \def\eoe{\eqno{\diamondsuit}}

\def\mapright#1{\smash{\mathop{\longrightarrow}\limits^{#1}}}
\def\mapdown#1{\Big\downarrow\rlap{$\vcenter{\hbox{$\scriptstyle#1$}}$}}

\def\mapse#1{\smash{\mathop{\searrow}\limits^{#1}}}

\def\interno{\hskip 2pt \vbox{\hbox{\vbox to .18
truecm{\vfill\hbox to .25 truecm
{\hfill\hfill}\vfill}\vrule}\hrule}\hskip 2 pt}

\begin{document}

\title{Asymptotic symmetry and asymptotic solutions to Ito stochastic differential equations}

\author{Giuseppe Gaeta$^{1,2}$, Roman
Kozlov$^{3}$, and Francesco Spadaro$^{4}$ \\
{ } \\
{ } \\
${}^{1}$ {\it Dipartimento di Matematica, Universit\`a degli Studi di Milano,} \\ 
{\it v. Saldini 50, 20133 Milano, Italy} \\
{ } \\ 
${}^{2}$ {\it SMRI, Santa Marinella, Italy} \\
{ } \\
${}^{3}$ {\it Department of Business and Management Science,} \\
{\it Norwegian School of Economics,} \\
{\it Helleveien 30, N-5045, Bergen, Norway} \\
{ } \\
${}^{4}$ {\it EPFL, CSFT, SB, Batiment MA - Station 8,} \\
{\it CH-1015 Lausanne, Switzerland}
}

\date{{\tt \giorno}}

\maketitle

\begin{abstract}
We consider several aspects of conjugating symmetry
methods, including the method of invariants, with an
asymptotic approach. In particular we consider how to
extend to the stochastic setting several ideas which are well
established in the deterministic one, such as conditional, 
partial and asymptotic symmetries. A number of explicit examples are presented.

\end{abstract}


\newpage

\tableofcontents

\newpage

\section{Introduction}

The theory of \emph{symmetry} -- i.e., what is now known as Lie theory
-- was developed by Sophus Lie in his attempt to solve (nonlinear)
differential equations, generalizing the approach by Evariste
Galois for algebraic equations.

The Lie approach, and more generally symmetry methods for
differential equations, produced a number of results by Lie and
his pupils, but afterwards the theory remained at the same level
for quite a long time, until it was taken up by Birkhoff in the
USA and by Ovsjannikov in USSR in the sixties \gcite{Bir,Ovs}. One
of the reason for this ``long sleep'' lies in that the applications of
the Lie theory do in general require quite extensive computations;
these can nowadays be routinely performed on a computer by
algebraic manipulation programs, but were quite prohibitive in earlier times.

After the reviving by Birkhoff and Ovsjannikov, the theory
developed quickly, and underwent various generalizations; the reader can consult some of the by now standard texts \gcite{\symmref} for the basics as well as for the most common extensions and generalizations.

Here we are concerned with two of these extensions and
generalizations. That is, on the one side with the generalization
to \emph{conditional} \gcite{LeWi}, \emph{partial} \gcite{CGpart},
and \emph{asymptotic} \gcite{Gasy} symmetries; and on the other
side with the extension of the theory to the framework of
\emph{stochastic} differential equations \gcite{\stochsymmref}
(by this we always mean equations of Ito or Stratonovich type \gcite{\sderef}, see below).

Our goal is to put the two together, i.e., to consider
\emph{conditional and asymptotic symmetries for stochastic
differential equations}. Needless to say, our main interest is not
in the abstract definition of these, but in showing that they can
be of concrete use in determining (conditional or asymptotic)
solutions to the concerned equations.

In the first part of the paper we establish some background.
In particular, we quickly recall some of the main points of
the theory of symmetry for (deterministic) differential equations,
mainly to fix notation (Section \ref{sec:symm}), then recall the
(well-known) notion of conditional symmetry (Section \ref{sec:cond}) and of
(less well-known) asymptotic symmetry (Section \ref{sec:asymp}) together with their use. We will also recall the basics of the theory of \emph{symmetry for
stochastic differential equations}  (Section \ref{sec:symmsto}), and the \emph{method of invariants} for these (Section \ref{sec:invar}).

We then proceed to merge the two, i.e., we develop \emph{conditional
and asymptotic symmetries and invariants for stochastic differential equations};
this will be the subject of Sections \ref{sec:cassto} and \ref{sec:caisto}. We also provide, in Section \ref{sec:examples}, a number of concrete
examples, chosen to be computationally simple in order to keep the conceptual issues in focus.

Our main result can be shortly stated saying that {\it the theory of conditional, partial and asymptotic symmetries can be extended to the framework of stochastic differential equations}. The reader will not find ``big theorems'' in our discussion, rather a number of lemmas; this corresponds to the fact we are not overcoming any special technical difficulty here. We are instead proposing a point of view that has not been considered previously in the literature, i.e., building a method which -- we believe -- is useful in characterizing solutions to stochastic equations via their asymptotic properties, in the same way as the corresponding deterministic approach proved to be precious in analyzing solutions to deterministic equations having certain asymptotic properties.

As we will discuss in more detail in the following (see Sections \ref{sec:cond} and \ref{sec:asymp}), the key observation triggering the present approach is that while in several physical cases one has equations which are invariant -- say, for the sake of concreteness, under a rotation or scaling transformation -- and correspondingly for these there is a special set of solutions displaying such symmetry, it is of more general interest the case where either $(i)$ the solutions do have such symmetry only asymptotically in time and/or space (think e.g., to the case where the rotationally invariant solutions are stable, so that solutions which are \emph{not} rotationally invariant at $t=0$ but which are sufficiently near to those will, in the long run, acquire the rotational symmetry); or $(ii)$ the equations do \emph{not} have such a symmetry, but there are solutions which, at least asymptotically in time and/or space, display such a symmetry.

The trivial, but not meaningless example, is that of the zero solution, which surely has a very high degree of symmetry, and which can be a valid -- and maybe also stable or even attractive -- solution for equations with no symmetry.

A less extreme, more concrete and also physically relevant, example is provided, in the deterministic framework, by the Boussinesq equation of Fluid Dynamics: in this case a number of physically relevant -- and routinely observed in experiments -- solutions do have symmetry properties which are not shared by the equation; and they can be analytically determined though the \emph{conditional symmetries} approach. Similarly, stable solutions to several types of reaction-diffusion equations (including the prototypical Fisher-Kolmogorov equation) can be analytically determined through the method of \emph{asymptotic symmetries}.

It must be stressed that in general, in this way one is able to obtain exact analytical expression for these symmetric class of solutions albeit there is no way to obtain a general solution to the equation under study in analytical terms.
We will thus extend these approaches, developed in the framework of deterministic equations; and propose a general method to study special classes (characterized indeed by symmetry-like properties) of solutions to \emph{stochastic} differential equations.

As already mentioned, we will illustrate the method by a number of concrete examples; these will however be chosen to be computationally simple in order to keep the conceptual issues in focus.

Throughout the paper, we will use the Einstein summation convention. We  will denote by $t$ the time variable and by $x^i$ the spatial ones, while the  $w^k$ will be Wiener processes; we moreover use the shorthand notation
$$ \pa_t \ := \ \frac{\pa}{\pa t} \ , \ \ \pa_i \ := \ \frac{\pa}{\pa x^i} \ , \ \ \^\pa_k \ := \ \frac{\pa}{\pa w^k} \ , $$
and denote the time derivative of a function $x (t)$ as either $x_t$ or $\xd$, depending on typographical convenience.

\section{Symmetry of differential equations}
\label{sec:symm}

In this section we recall some basic notions concerning
symmetries of deterministic differential equations. In particular
we consider first order ODEs, i.e., dynamical systems (DS), as
this is the deterministic analogue of Ito or Stratonovich
equations to be considered in the following.

For a dynamical system \beql{eq:DS} \frac{d x^i}{d t} \ = \ f^i
(x,t) \eeq with $t \in R$, $ x \in M_0 \sse R^n$, we define $M =
M_0 \times R$ and consider maps $\Phi : M \to M$. In particular,
we consider one-parameter groups of such maps, with generator
\beql{eq:X} X \ = \ \theta (x,t) \, \pa_t \ + \ \vphi^i (x,t) \,
\pa_i \ . \eeq

Note that with $\pi_0 : M \to R$ the projection $\pi_0 (m_0,r) = r$, the full phase bundle $M = M_0 \times R$ has a natural
structure of fiber bundle over the time real axis, $(M,\pi_0,R)$.

Under the action of $X$ -- or more precisely of the maps $\Phi =
\exp[ \eps X]$ in the one-parameter group generated by $X$ -- we
have, at first order in $\eps$, \beq t \to \wt{t} \ = \ t \ + \
\eps \, \theta (x,t) \ , \ \ x^i \to \wt{x}^i \ = \ x^i \ + \ \eps
\, \vphi^i(x,t) \ . \eeq Once we define the transformation of the
independent ($t$) and the dependent ($x^i$) variables, we have
also implicitly defined the transformation of the derivatives of
the latter with respect to the former. We have indeed (always working at first order in $\eps$)
$$ x^i_t \ \to \ \wt{x}^i_t \ = \ x^i_t \ + \ \eps \psi^i (x,t,\xd) $$
where $\psi^i$ is given by the \emph{prolongation formula}
\gcite{\symmref} \beql{eq:psi} \psi^i (x,t,\xd) \ = \ D_t \vphi^i \
- \ x^i_t \ D_t \theta \ ,  \eeq and $D_t$ is the total time
derivative (in this case, this is just $D_t = \pa_t + \xd^j
\pa_j$).

With this, the vector field $X$ on $M$ is extended (prolonged) to
a vector field $X^{(1)}$ in $J^1 M$ (in which we consider local
coordinates $\{t;x^1,...,x^n;\xd^1,...,\xd^n\}$), \beql{eq:X1}
X^{(1)} \ = \ \theta (x,t) \, \frac{\pa}{\pa t} \ + \ \vphi^i
(x,t) \, \frac{\pa}{\pa x^i} \ + \ \psi^i (x,t;\xd) \,
\frac{\pa}{\pa \xd^i} \ . \eeq

Here and below $J^1 M$ is the (first) \emph{jet bundle} associated
to $M$ \gcite{\symmref}. Note this can be seen both as a bundle
over $M$ (the fiber corresponding to first derivatives $\xd$) with
projection $\wt{\pi}_1$ and as a bundle over $R$ (the fiber
corresponding to variables $x$ and first derivatives $\xd$) with
projection $\pi_1$,
$$
\matrix{ J^1 M & \mapright{\wt{\pi}_1} & M \cr
 & & \\
 & \mapse{\pi_1} & \mapdown{\pi_0}  \cr
 & & M_0 \cr} \ . $$
The projections satisfy $ \pi_1 =  \pi_0 \circ \wt{\pi}_1$, i.e.
the diagram above is a commutative one.

The DS \eqref{eq:DS} defines a submanifold in $J^1 M$, also called
the \emph{solution manifold} $\Sigma_f$ for the Eq
\eqref{eq:DS}.

A function $x = \xi (t)$ is identified with its graph $$ \ga_\xi \
= \ \{ (x,t) \ : \ x = \xi (t) \} \ \ss \ M \ , $$ i.e., with a
section $\ga_\xi$ of the bundle $(M,\pi_0,R)$; this is naturally
extended to a section $\ga_\xi^{(1)}$ in $J^1 M$. In local
coordinates, we have \beq \ga^{(1)}_\xi \ = \ \{ t ; \xi^1 (t) ,
... , \xi^n (t) ; \dot{\xi}^1 (t) , ... , \dot{\xi}^n (t) \} \ .
\eeq The function $x = \xi (t)$ is a solution to \eqref{eq:DS} if
and only if \beq \ga_\xi^{(1)} \ \ss \ \Sigma_f \ . \eeq

The vector field \eqref{eq:X} is a Lie-point symmetry (generator)
for the Eq \eqref{eq:DS} if and only if \beql{eq:Xsymm}
X^{(1)} \ : \ \Sigma_f \ \to \ \T \, \Sigma_f \ ; \eeq in this
case it also maps (sections of $(M,\pi_0,R)$ representing)
solutions to \eqref{eq:DS} into  -- generally, but not
necessarily, different -- (sections of $(M,\pi_0,R)$ representing)
solutions to \eqref{eq:DS} \gcite{\symmref}.

Note that we will adhere to the usual abuse of notation consisting in calling
``symmetries'' the vector fields which are actually generators for the symmetry group.

It may be worth noting that the action of the vector field \eqref{eq:X} on sections is given by
\beql{eq:Xsections} X \ : \ f(t) \ \to \wt{f} (t) \ = \ f (t) \ + \ \eps \ \[ \vphi \ - \ \theta \, f' (t) \] \ . \eeq
To avoid any misunderstanding, we specify that this notation means that the section $\Ga_f$ given by $\Ga_f = \{ (t,x=f(t))\}$ is mapped into the section $\Ga_{\wt{f}}$ given by $\Ga_{\wt{f}} = \{ (t,x=\wt{f} (t))\}$.

Actually this is also an equivalent characterization of
symmetries: that is, symmetries can be defined as maps $M \to M$
which transform solutions into (generally, but not necessarily,
different) solutions \gcite{\symmref}.\footnote{For a very readable discussion of different -- but equivalent -- definitions of symmetries, see e.g., Chapter 3 in ref.\gcite{Stephani} or Section 2.2. in ref.\gcite{Olver1}.}

If $x =\xi (t)$ is a solution to \eqref{eq:DS} and if $X$, satisfying
\eqref{eq:Xsymm}, leaves $\ga_\xi$ invariant, we say that $x =\xi
(t)$ is an \emph{$X$-invariant solution} to \eqref{eq:DS}.

Vector fields that are symmetries for a given equations, such as the evolution Eq \eqref{eq:DS}, or of more general form,
-- or for equations, be these ODEs or PDEs, of more
general form -- are characterized as solutions to a set of
\emph{linear} PDEs for the coefficients of the vector field, known
as \emph{determining equations} \gcite{\symmref}. For equations of
the form \eqref{eq:DS} and vector fields expressed as in
\eqref{eq:X} with $\theta = 0$ (these are the simple vector fields to be considered below), these are written as \beql{eq:deteqdet} \frac{\pa
\vphi^i}{\pa t} \ + \ f^j \, \frac{\pa \vphi^i}{\pa x^j} \ - \
\vphi^j \, \frac{\pa f^i}{\pa x^j } \ = \ 0 \ . \eeq

\medskip\noindent
{\bf Remark \ref{sec:symm}.1.} It should be mentioned that this sketchy discussion refers to proper Lie-point symmetries. In the case of deterministic Dynamical Systems, it turns out that so-called \emph{orbital symmetries} are also quite useful \gcite{WalOS1,WalOS2,WalOS3}; as far as we know, these have not been investigated in the context of \emph{stochastic} dynamical systems. \EOR

\medskip\noindent
{\bf Remark \ref{sec:symm}.2.} The prolongation formula for ODEs, Eq \eqref{eq:psi}, is a special occurrence of the general
prolongation formula, which -- in the multi-index notation used in most of the literature, and in particular denoting by $x^i$ the independent variables, and by $u^a$ the dependent ones -- reads
$$ \psi^a_{J,i} \ = \ D_i \vphi^a_J \ - \ u^a_{J,k} \ D_i \xi^k \ . $$
We refer to \gcite{\symmref} for discussion and details. \EOR

\subsection{Dynamically invariant submanifolds}

The evolution Eq \eqref{eq:DS} defines a dynamics, and a
dynamical vector field $Z$, in all of $M$; when $f$ is time-autonomous, we actually also have a vector field $Y$ in $M_0$ and
in this case $Z = \pa_t + Y$; symmetries $X$ are then characterized by $[X,Z] = 0$ (so called \emph{orbital symmetries} are characterized by $[X,Z] = \la Z$, for their properties and applications see e.g., \gcite{WalOS1,WalOS2,WalOS3}).  We are specially interested in this latter case of $f$ time-autonomous, both because the analysis is slightly simpler and due to its relevance in
applications.

It may happen that there exist submanifolds of $M$ being invariant
under $Z$. In particular, for $f$ time autonomous, there may be
proper submanifolds $U \ss M_0$ which are invariant under $Y$. The
simplest example is that of fixed points for $f$, i.e., points
$x_0$ such that $f(x_0,t) = 0$. Other simple but significant
examples are provided by periodic or multi-periodic solutions, by
heteroclinic or homoclinic orbits, or by the stable and unstable
manifolds of hyperbolic fixed points \gcite{\dsref}.

One would expect that symmetries are constrained by the need to
somehow respect the structure of such invariant sets; this is indeed
the case, provided one considers time-independent symmetries \gcite{CGbook,GLPTI}.
In particular, it can be shown that Lie-point time independent (LPTI)
symmetries maps solutions into solutions whose trajectories in $M_0$
have the same topology.

More specifically, we have the following results (these embody
several Lemmas given in \gcite{CGbook,GLPTI}; see there for proofs; the Corollary makes use of standard results in Topology, see e.g., \gcite{Milnor}). We
will denote by $K(X) \sse M$ the subset of $M$ which is point-wise
invariant under a vector field $X$ (or $K(\mathcal{A}) \sse M$ for the set invariant under an algebra $\mathcal{A}$ of vector fields), by $Z$ the vector field defining the dynamics of the (autonomous) dynamical system under study, and by $\mathcal{G}$
the Lie algebra of symmetries for $Z$.

\medskip\noindent
{\bf Lemma \ref{sec:symm}.1.} {\it The LPTI symmetry algebra of an
autonomous ODE transforms stationary solutions into stationary
solutions, periodic solutions into periodic solutions of the same
period.}

\medskip\noindent
{\bf Lemma \ref{sec:symm}.2.} {\it Assume there is a compact set $K
(Z) \ss M$ invariant under the flow of $Z$. Then for any
subalgebra $\G_0 \sse \G$, the submanifold $K^0 (\G_0 ) := \(
K(\G_0) \cap K (Z) \) \sse M$ is either empty or compact and
invariant under the flow of $Z$.}

\medskip\noindent
{\bf Corollary.} {\it If $K^0 (\G_0 )$ has components isomorphic
to an even-dimensional sphere $S^{2n}$, then on each of these lie
stationary solutions to the dynamical system $\xd = Z (x)$. If
$K^0 (\G_0 )$ has components isomorphic to a disk $D^{2n+1}$, then
on each of these lie stationary solutions to $\xd = Z
(x)$.}

\subsection{Asymptotic solutions}

In several cases one meets dynamical systems or PDEs which cannot
be solved exactly, but whose asymptotic behavior (e.g., for $t \to
\infty$ and/or for $|x| \to \infty $ or for $|x|\to 0$) is
amenable to exact treatment.

We just mention here two very simple examples of use in the
following (these can be readily generalized).

\medskip\noindent
{\bf Example \ref{sec:symm}.1.}
For $x \in R$, consider
$$ \xd \ = \ f(x) \ = \ - \nabla \Phi (x) $$
with $\Phi (x) = \Phi (-x)$ a convex even function, non-degenerate in $x=0$. Then for any
initial datum $x(0) =x_0$ the dynamics is attracted to $x_* = 0$,
and in the region near $x_*$ (and thus asymptotically in time) the
dynamics is well described by the linear equation
$$ \xd \ = \ - \, A \, x \ := \ [(D f)(0)] \, x \ , \ \ \ A = - (Df)(0) > 0 \ ; $$
the solutions to this are of course
$$ x(t) \ = \ x(t_0) \ \exp[ - A t] \ . \eoe $$

\medskip\noindent
{\bf Example \ref{sec:symm}.2.} In $R^2$ and with polar
coordinates $(\rho,\vartheta)$, consider the system
\begin{eqnarray*}
\dot{\rho} &=& \rho \ \( 1 \ - \ \rho \) \ , \\
\dot{\vartheta} &=& 1 \ + \ \( 1 \ - \ \rho \) \ h (\rho,\vartheta
) \ , \end{eqnarray*} with $h (\rho,\vartheta )$ an arbitrary
smooth function. The system can not be solved in general, but we
know that the dynamics is attracted to the circle $\rho=1$, and on
this attracting manifold we have a trivial dynamics
$\dot{\vartheta} = 1$. Thus any solution is asymptotically
described by $\rho=1$, $\vartheta (t) = k_0 + t$ for some phase
$k_0$. \EOE

\section{Conditional and partial symmetries}
\label{sec:cond}

As mentioned above, symmetries map (any) solution into a
(generally different) solution. It may happen, however, that there
are maps such that this property (i.e., mapping solutions into
solutions) holds only for a \emph{subset} of solutions.

\subsection{Conditional symmetries}

The most common case is that of equations admitting solutions
which are invariant under maps that are \emph{not} a symmetry of
the equation. For instance, any equation of the form \eqref{eq:DS} with
$f$ linear in $x$, $f(x,t) = A(t) x$,  will admit the origin as a
fixed point, and this in turn will be invariant under any linear
(possibly time-dependent) map $x \to B(t) x$. But this will be a
symmetry of the equation if and only if
$$ \frac{d B (t)}{d t} \ = \ \left[ A (t) \, , \, B(t) \right]  \ , $$
which is definitely a non-generic property.

In this case, and in similar but less trivial ones, one speaks
also of \emph{conditional} symmetries \gcite{LeWi}; see also
\gcite{BluCol,BluColB,ClaKru,ClaMan,Gclw,PuS1,PuSn,Zhda}. Let us discuss
these in some more detail (the reader can see the references given
above for a more complete discussion), referring to the case of
PDEs for greater generality; we denote by $x^1,..,x^q$ the
independent variables, and by $u^1,...,u^p$ the dependent ones.
Our goal is to determine $X$-invariant solutions for a
differential equation (note that by this we will always mean
possibly a vector one, i.e., a system of scalar differential equations)
$\De = 0$, where
$$ X \ = \ \xi^i (x,u) \ \frac{\pa}{\pa x^i} \ + \
\vphi^a (x,u) \ \frac{\pa}{\pa u^a} \ = \ \xi^i \,\pa_i \ + \
\vphi^a \, \pa_a $$ is not necessarily a symmetry of $\De$; at the
same time, we want to determine the $X$ which admit such
solutions.

In order to do this, we complement $\De$ with an equation $\De_X=
0$ expressing the requirement of $X$-invariance of the function $u
= u(x)$. This is just, see \eqref{eq:Xsections} above, \beql{eq:Xinv}
\vphi^a (x,u) \ - \ \xi^i (x,u) \, u^a_i \ = \ 0 \ . \eeq We are
thus led to consider the system \beql{eq:Ecs} \mathcal{E}_X \ = \
\cases{ \De \ = \ 0 & \cr \De_X \ = \ 0 & \cr} \eeq It is clear
that the solutions of this, \emph{if they exist}, are all and only
the solutions to $\De=0$ which are also $X$-invariant (i.e., which
also solve $\De_X= 0$). Thus symmetries for the system
$\mathcal{E}_X$, which transform solutions into solutions, will
transform $X$-invariant solutions to $\De$ into $X$-invariant
solutions to $\De$. In particular this apply for the vector field
$X$ itself, which by construction will leave each of such
solutions invariant.

We are thus at first sight reduced to a standard problem in
symmetry analysis; however we should look at this with some more
care. One possibility is that we fix in advance $X$ and look for
solutions to $\De$ which are $X$-invariant. In this case $\De_X$
reduces to an \emph{ansatz} on the functional form of $u(x)$ which
can then be inserted into $\De$, and indeed nothing new is here.

On the other hand, the interesting case is the one where we do not
know in advance what $X$ should or could be, and we study the
system \eqref{eq:Ecs} in order to determine $X$ and the associated
invariant solutions, if any.

In this case we get something new: in fact, the usual determining
Eq \eqref{eq:deteqdet} -- or more generally their version for PDEs -- are \emph{linear} in the $X$ coefficients
$\{ \xi^i (x,u) , \vphi^a (x,u) \}$, while now we will have
determining equations which are \emph{nonlinear} (in particular,
quadratic) in these.

We will not enter into the details of how such a system of
nonlinear (determining) equations can be solved thanks to the
peculiar structure of the problem, referring the reader to the
original paper by Levi and Winternitz for this matter \gcite{LeWi}.

We instead stress a most relevant point: symmetries have to map
solutions into solutions, but when we apply symmetry analysis to
the system \eqref{eq:Ecs}, this applies to the \emph{common}
solutions to $\De=0$ and to $\De_X = 0$, i.e., \emph{only} to
invariant solutions. Thus these are symmetries of $\De$
\emph{conditional} upon the additional invariance condition
$\De_X$.

\medskip\noindent
{\bf Remark \ref{sec:cond}.1.} We should mention that one could also investigate the inverse problem for conditional symmetries: that is, for a given number of dependent and independent variables and a given order of the equation, and given a certain vector field $X$, determine all the differential equations $\De$ which admit $X$ as a conditional symmetry. This was considered by Levi, Rodriguez and  Thomova in \gcite{LRT}, and more recently by Pucci and Saccomandi \gcite{PS1,PS2}. We are not able to provide stochastic extensions of this analysis at the moment. \EOR

\subsection{Partial symmetries}

A slightly more general case is also possible, i.e., that we have a
subset of solutions which are \emph{not} individually invariant
under the map, but which are indeed transformed one into another
by the map -- which again is not a symmetry of the original
equation $\De = 0$. In this case one speaks of \emph{partial
symmetries} \gcite{CGpart} (see also \gcite{GaKa}).

A specially relevant case in our dynamical systems context is 
where the
solutions in this set are characterized by living on a manifold which is dynamically invariant (i.e., invariant under the dynamics).
The situation is characterized in a way which looks quite similar
to the characterization provided above for conditional symmetries.
That is, we consider a differential equation (again, in general
this can be a system of PDEs) $\De = \De_0$, and a vector field
$X$ which is \emph{not} a symmetry for $\De_0$, thus such that --
writing $S_0 :=S_{\De_0}$ for ease of writing,
$$ \[ X^{(n)} \, \De_0 \]_{S_0} \not= \ 0 \ . $$
We consider then an auxiliary equation
$$ \De_1 \ := \ \[ X^{(n)} \, \( \De_0 \) \] \ . $$
If we look at solutions to the system
$$ \mathcal{E} \ = \ \cases{ \De_0 \ = \ 0 & \cr \De_1 \ = \ 0 & \cr} $$
which are globally invariant under $X$ (that is, the set of
solutions is mapped into itself, albeit the single solutions are
possibly not invariant), this set is characterized precisely by
the property \beql{eq:part1} \[ X^{(*)} \( \De_1 \) \]_{S^{(1)}} \
= \ 0 \ ; \eeq here $S^{(1)} = S_0 \cap S_1$, with of course $S_1
=S_{\De_1}$, is the set of solutions to the system $\mathcal{E}$;
and we wrote $X^{(*)}$ to mean the prolongation of $X$ of
appropriate order, i.e., the same order as the equation $\De_1$. In
this case, i.e., if \eqref{eq:part1} is satisfied, we say that
$X^{(*)}$ is a partial symmetry of order one for $\De$, as it maps
a subset of solutions into solutions (belonging to the same
subset).

We stress that \eqref{eq:part1} may or may not be satisfied,
depending on $\De$ and on our choice of the vector field $X$. If
it is not satisfied, we can iterate our procedure and consider
$$ \De_2 \ := \ X^{(*)} \( \De_1 \) $$
and look for $S^{(2)} = S_0 \cap S_1 \cap S_2$, enquiring if --
with an obvious notation -- the equation \beql{eq:part2} \[
X^{(*)} \( \De_2 \) \]_{S^{(2)}} \ = \ 0  \eeq is satisfied. If
this is the case, we say that $X$ is a partial symmetry of order
two for $\De$; otherwise we can still iterate our procedure and so
on.

We set this in the form of a Proposition, which is quoted from
\gcite{CGpart}.

\medskip\noindent
{\bf Proposition \ref{sec:cond}.1.} {\it Given the general
differential problem $\De = \De_0 = 0$ and a vector field $X$ on
$M$, define $\De_{r+1} := X^{(*)} (\De_r)$. Denote by $S^{(r)}$
the set of simultaneous solutions of the system
$$ \De_0 \ = \ \De_1 \ = \ ... \ = \ \De_r \ = \ 0 \ , $$
and assume that this is not empty for $r \le s$. Assume moreover that
$$ \[ X^{(*)} \( \De_r \) \]_{S^{(r)}} \ \not= \ 0 \ \ \mathrm{for} \ r = 0, 1, ... , n-1 \ , \ \ \
\[ X^{(*)} \( \De_n \) \]_{S^{(n)}} \ = \ 0 \ . $$
Then the set $S^{(n)}$ provides a family of solutions to the
initial problem $\De$ which is mapped into itself by the
transformations generated by $X$.}

\medskip\noindent
{\bf Remark \ref{sec:cond}.2.} When the situation depicted in
Proposition \ref{sec:cond}.1 is met, we shall say that $X$ is a
\emph{partial symmetry}, or P-symmetry for short, for the problem
$\De$, and that the globally invariant subset of solutions
$S^{(n)}$ obtained in this way is a \emph{$X$-symmetric set}. We
also refer to the number $n$ appearing in the statement as the
\emph{order} of the P-symmetry. \EOR

The discussion can be specialized to the case of dynamical system;
here we  just recall a result for this case, quoting again
from \gcite{CGpart}, to which the reader is referred for further
detail. Here $\L_{(\la)} (f)$ denotes the linearization of $f$ at
$x_{(\la)}$ (see below), i.e.,
$$ \L_{(\la)} (f) \ = \ \[ \nabla  ( f ) \]_{x_{(\la)}} \ . $$

\medskip\noindent
{\bf Proposition \ref{sec:cond}.2.} {\it Assume that the dynamical
system $\xd = f(x)$ admits a LPTI partial symmetry $X= \vphi^i(x)
\pa_i$, and let $x_{(\la)} = x_{(\la)} (t) $ be an orbit of
solutions obtained under the action of the group generated by $X$,
with $\la$ the group parameter. Then $\Phi := \vphi (x_{(\la)} )$
satisfies the equations
$$ \frac{d \Phi}{d t} \ = \ \L_{(\la)} (f) \cdot \Phi  \ ; \ \ \
\Phi \ = \ \frac{d x_{(\la)}}{d \la} \ . $$}

\subsection{Examples}
\label{sec:condpartexa}

We give here two Examples (examples \ref{sec:cond}.1 and \ref{sec:cond}.2) of conditional symmetries, and three Examples (examples \ref{sec:cond}.3 through \ref{sec:cond}.5) dealing with partial symmetries. These are taken respectively from \gcite{LeWi} and from \gcite{CGpart}. Some Examples deal with PDEs, as the essential of conditional and partial symmetries is better grasped in this context.

\medskip\noindent
{\bf Example \ref{sec:cond}.1.} The Boussinesq equation \gcite{Whitham}
in one spatial dimension, for a function $u =u(x,t)$, reads
\beql{eq:Bou} u_{tt} \ + \ u \, u_{xx} \ + \ u_x^2 \ + \ u_{xxxx}
\ = \ 0 \ . \eeq If we consider the vector field
$$ X_1 \ = \ \pa_t \ + \ t \, \pa_x \ - \ 2 \, t \, \pa_u \ , $$
this is not a symmetry for the equation, but it turns out to be a
conditional symmetry for it.

Under the action of $X_1$ we have a one-parameter flow (group
parameter $\la$) which acts as
$$ \cases{ t \ \to \ \wt{t} \ = \ t + \la & \cr
x \ \to \ \wt{x} \ = \ x + \la t + (1/2) \la^2 & \cr u \ \to \
\wt{u} \ = \ u - 2 t \la - \la^2 & \cr} $$ Assuming that $u=u(x,t)$
is a solution to \eqref{eq:Bou}, we have that $\wt{u} = \wt{u}
(\wt{x}, \wt{t})$ is a solution if it satisfies the auxiliary
condition (we omit the tildes from now on) \beql{eq:BouLW1} \De_1
\ := \ \( u_t \ + \ t \, u_x \)_x \ = \ 0 \ . \eeq

Now we note that the invariance condition $\De_X = 0$, see
\eqref{eq:Xinv}, reads in this case (i.e., for the $X$ we are now
considering)
$$ u_t \ + \ t \, u_x \ + \ 2 \, t \ = \ 0 \ . $$
Thus \eqref{eq:BouLW1} is just a differential consequence of this,
$(\De_1)_x = 0$. \EOE

\medskip\noindent
{\bf Example \ref{sec:cond}.2.} Consider again the Boussinesq
Eq \eqref{eq:Bou}, and now
$$ X_2 \ = \ \pa_t \ - \ (x/t) \, \pa_x \ + \ \[ (2/t) \, u \ + \
(6/t^3) \, x^2 \] \, \pa_u \ ; $$ again this is \emph{not} a
symmetry for Eq \eqref{eq:Bou}. Now we get
$$ \wt{u} \( \wt{x} , \wt{t} \) \ = \ \[ \frac{t +\la}{t} \]^2 \ u(x,t)
\ + \ \[ \frac{(t+\la)^2}{t^4} \ - \ \frac{t^2}{(t+\la)^4} \] \ x^2 \ . $$
On the other hand, the invariance condition \eqref{eq:Xinv} is now
\beql{eq:BouLW2} \De_2 \ := \  \[ \frac{2}{t} \, u \ + \ \frac{6}{t^3} \, x^2 \]
\ - \ u_t \ + \ \frac{x}{t}  \, u_x \ = \ 0 \ . \eeq

If $u(x,t)$ is a solution to \eqref{eq:Bou}, then the condition
for $\wt{u} ( \wt{x} , \wt{t} )$ to be also a solution reads
$$A_1 (x,t;\la) \ F \ + \ A_2 (x,t;\la) \ F_t \ + \ A_3 (x,t;\la) \ F_x \ = \ 0 \ , $$
where $A_i$ are certain smooth polynomial functions, and
$$ F \ := \ u_t \ - \ (x/t) \, u_x \ - \ \[ (2/t) \, u \ + \ 6 \, (x^2/t^3) \] \ . $$
Thus if $u(x,t)$ is a solution to the Boussinesq equation and
$F=0$, the transformed function $\wt{u} ( \wt{x} , \wt{t} )$ is
also a solution. But looking at \eqref{eq:BouLW2} we see that
$F=0$ is just the invariance condition under $X_2$. \EOE

\medskip\noindent
{\bf Example \ref{sec:cond}.3.} Consider the modified Laplace
equation \beql{eq:MLE} \De \ := \ \ u_{xx} \ + \ u_{yy} \ + \ g(u) \,
u_{xxx} \ = \ 0 \ , \eeq with $g(u)$ an arbitrary smooth function.
We also consider the rotation vector field \beql{eq:MLEX} X \ = \ y \, \pa_x
\ - \ x \, \pa_y \ ; \eeq this is a symmetry for $\De$ in the case
$g(u)=0$, but not in general. We will assume $g(u) \not= 0$.

Implementing our procedure, we get
\begin{eqnarray*}
\De_1 & := & \ u_{xxy} \ , \\
\De_2 & := & \ 2 \, u_{xxy} \ - \ u_{xxx} \ , \\
\De_3 & := & \ 2 \, u_{yyy} \ - \ 7 \, u_{xxy} \ , \\
\De_4 & := & \ 3 \, u_{xyy} \ , \\
\De_5 & := & \ u_{yyy} \ - \ 2 \, u_{xxy} \ . \end{eqnarray*}

The last equation is identically satisfied on common solutions to
the previous ones, thus we have a partial symmetry of order five.

The set $S^{(5)}$ has the following form, with $k_{1},\dots,k_{6}$ arbitrary constants:
$$ S^{(5)} \ = \ \{ u(x,y) \ = \ k_1 \, x^2 \ + \ k_2 \, y^2 \ + \ k_3 \, x y \ + \
k_4 \,x \ + \ k_5 \, y \ + \ k_6  \} \ ; $$
that is, functions which are at most quadratic in the $(x,y)$ variables. Note that for these functions the Eq \eqref{eq:MLE} reduces to the standard Laplace equation, and for this it is obvious that \eqref{eq:MLEX} is a symmetry; thus this seemingly complicated computation yields a rather obvious result. \EOE

\medskip\noindent
{\bf Example \ref{sec:cond}.4.} We  look again at the
Boussinesq Eq \eqref{eq:Bou}, which will be our $\De_0$,
and consider the vector field \beq X \ = \ \pa_t \ + \ t \,\pa_x \
- \ 2 \, t \, \pa_u \ . \eeq Now we have (disregarding some inessential nonzero multiplicative constants)
\begin{eqnarray*}
\De_1 &=& X^{(*)} \, \De_0 \ = \ u_{xt} \ + \ t \, u_{xx} \ ,\\
\De_2 &=& X^{(*)} \, \De_1 \ = \ 0 \ . \end{eqnarray*}

We have thus to look for simultaneous solutions to $\De_0=0$ and
to $\De_1=0$. Solving for $\De_1=0$ yields \beql{eq:partB1} u(x,t)
\ = \ v (x -t^2/2) \ + \ w(t) \ - \ t^2 , \eeq with $v$ and $w$
arbitrary functions. Note that the functions \eqref{eq:partB1} are
$X$-invariant if and only if $w'(t)=0$ (this is why the
term $- t^2$ has not been absorbed into the arbitrary function
$w(t)$ in writing \eqref{eq:partB1}); thus we are
considering a setting which is really a generalization of the
conditional symmetries approach, even in this case.

Inserting now \eqref{eq:partB1} into $\De_0$, i.e. into the
Boussinesq equation \eqref{eq:Bou}, and writing $z = x -t^2/2$, we
get that $v$ and $w$ must satisfy \beq \frac{d}{dz} \( v''' \ + \
v \, v' \ - \ v \ - \ 2 \, z \) \ + \ w \, v'' \ + \ \frac{d^2
w}{d t^2} \ = \ 0 \ . \eeq We are specially interested, as
mentioned above, in the case $w (t)\not= 0$ (otherwise we recover
the Levi-Winternitz family of $X$-invariant solutions). E.g., for
$w(t) = K$ a nonzero constant, we get solutions $u(x,t)= K+v(z)-t^2$
with $v$ a solution to
$$ v''' \ + \ v \, v' \ - \ v \ + \ K \, v' \ = \ c \ + \ 2 \, z \ . \eoe $$

\medskip\noindent
{\bf Example \ref{sec:cond}.5.} We now consider an example
involving a Dynamical System in $R^3$ with Cartesian coordinates
$(x,y,z)$; we will also write $\rho^2 = x^2 + y^2$. Consider
$$ \cases{
\dot{x} \ = \ x \, (1 -\rho^2) \ - \ y \ + \ z \, g_1 (x,y,z) & \cr
\dot{y} \ = \ y \, (1 -\rho^2) \ + \ x \ + \ z \, g_2 (x,y,z) & \cr
\dot{z} \ = \ z \, g_3 (x,y,z) & \cr} $$
with $g_i$ smooth functions. If we also consider the rotation vector field
$$ X \ = \ y \, \pa_x \ - \ x \, \pa_y \ , $$
then the partial symmetry condition reads
$$ z \ G_\a (x,y,z) \ = \ 0  \ \ \ (\a = 1,2,3) $$
with $G_\a$ certain functions which are non-zero for generic
$g_i$: e.g., $$ G_1 \ = \ g_2 \ - \ y \, (\pa g_1 / \pa x) \ + \ x
\, (\pa g_1 / \pa x) \ . $$

Thus the system exhibits -- as rather obvious from its expression
-- rotational $(x,y)$ symmetry only if restricted to the plane
$z=0$. Moreover, in this plane one can find three different
families of solutions (beside the null one, which is rotationally
invariant) which are mapped into themselves by the rotations,
without being rotationally invariant themselves: the trajectories
lying in $\rho < 1$, and respectively in $\rho > 1$, spiralling
towards the limit cycle $\rho = 1$; and the solutions running on
the single trajectory $\rho =1$ which is left fixed by the partial
symmetry (i.e., the limit cycle). \EOE

\section{Asymptotic symmetries}
\label{sec:asymp}

In many cases, one has equations whose solutions have a complicated
behavior, which becomes simple in some asymptotic regime. This may
be asymptotic in time (thus for $t \to \pm \infty$, possibly with
different behavior in the two limits or with a simple behavior
only in one of the limits) or asymptotically in space (i.e., for
$|x| \to \infty$, or in the one dimensional case for $x \to \pm
\infty$, with the same notes as for $t$); or in a suitable
combination of the two.

It is remarkable that in such cases one can adopt a
renormalization-group (RG) like approach \gcite{GO,BK}; this in
turn can be combined with symmetry considerations
\gcite{Gasy,GMan1,GMan2}. In particular, we can have \emph{asymptotic
symmetries}, i.e., maps which are symmetries (hence in particular
map solutions into solutions) only in this asymptotic regime,
being otherwise (that is, in the non-asymptotic regime) only
approximate symmetries, at most.

\subsection{Asymptotic symmetries in the PDE setting}

In order to discuss asymptotic symmetry properties (at least in
the deterministic case) we consider the general case, i.e.,
possibly PDEs (of any order $n$); the independent variables live
on a manifold $B$, the dependent ones take values in a manifold
$U$, and we consider $M = U \times B$. We will follow the
discussion given in \gcite{CGbook}, referring to \gcite{Gasy,GMan1,GMan2}
for further detail.

It is then convenient to use the geometrical representation of
functions as cross sections of the bundle  $(M,\pi_0,B)$, and of
equations as submanifolds in $J^n M$; in the case of evolution
equations, these can also be seen as sections of a suitable fiber
bundle (with time derivatives in the fiber, i.e., in the vertical
space), and on the other hand some care is needed in order to take
into account the special role of the time variable in this case.

Here and in the following we
write $\pa_a^J = (\pa/ \pa u^a_J)$; moreover $\psi^a_0 =
\vphi^a$, and $u^{[n]}$ will denote $(u, u^{(1)}, ... , u^{(n)} )$. Recall also that $|J|$ is the order of the multi-index $J$.

A vector field
\beq Y \ = \ \xi^i (x,u) \,\pa_i \ + \ \vphi^a (x,u) \, \pa_a \eeq
in $M$, and its prolongation \beq Y^{(n)} \ = \ \xi^i (x,u) \,\pa_i \ + \ \psi^a_J
(x,u^{[|J|]}) \, \pa_a^J \eeq to $J^n M$, will then induce an action on the
space of sections in $(M,\pi_0,B)$, hence on functions; and on
submanifolds in $J^n M$, hence on equations.

In particular, a (vector) function $u^a = f^a (x)$ is then mapped by the infinitesimal action of the vector field (with parameter $\eps$)
into \begin{eqnarray} \wt{u}^a (x) &=& \( e^{\eps Y} f \)^a (x) \nonumber \\
&=& u^a (x)
\ + \ \eps \ \[ \vphi^a (x,f(x)) \ - \ \xi^i (x,f(x)) \, \pa_i f^a
(x) \] \ + \ O (\eps^2 ) \ ; \end{eqnarray} this infinitesimal action defines a flow (with parameter $\la$, not necessarily small) $\ga (\la) =
\Phi(\la,\ga_0)$ in the set $\Ga (M)$ of sections in
$(M,\pi_0,B)$, where $\la$ is the flow parameter and $\ga_0$ is
the initial point for the flow. From this one also computes the
transformation rules for derivatives $u^a_J$, i.e., on sections of
$(J^n M,\pi_n, B)$, and hence for equations; this also defines a
flow $\ga^{(n)} (\la) = \Phi^{(n)} (\la,\ga^{(n)}_0)$ in the set
$\Ga^{(n)} (M)$ of sections in $(J^n M,\pi_n,B)$, where $\la$ is
the flow parameter and $\ga^{(n)}_0$ is the initial point for the
flow.

We will also write these flow as \beq f_\la = \Phi (\la,f_0) \ \ , \ \ \ \
\De_\la = \Phi^{(n)} (\la, \De_0) \ , \eeq for ease of writing.

We say that a function $u = f(x)$ is symmetric under $Y$ if and
only if the corresponding section $\ga_f \in \Ga (M)$ is invariant
under (the flow $\Phi$ in $\Ga$ induced by) $Y$; and similarly
that the equation $\De$ in $J^n M$ is symmetric under $Y$ if and
only if the corresponding submanifold $S_\De$ is invariant under
(the flow $\Phi^{(n)}$ in $J^n M$ induced by) the prolongation
$Y^{(n)}$.

We are specially interested in the case where the flow drives
functions (or equations) which are \emph{not} invariant to a
\emph{fixed point}, i.e., to a limit function (or an equation)
which \emph{is} invariant.

In particular, consider the case where the limit
\beq \lim_{\la \to \infty} \Phi (\la , \ga_f ) \ = \ \ga_f^\infty \eeq
exists. Then we say that $Y$ is an $\om$-asymptotic symmetry for
(the section $\ga_f$ and hence) the function $f:B \to
U$. (In the case where the limit $ \lim_{\la \to - \infty}
\Phi (\la , \ga_f ) \ = \ \ga_f^{-\infty} $ exists, we say that
$Y$ is an $\a$-asymptotic symmetry for $f$. This notation recalls
the notions of $\om$-limit and of $\a$-limit in use in Dynamical
Systems theory \gcite{\dsref}.)

Similarly, in the case where the limit
\beq \lim_{\la \to \infty} \Phi^{(n)} (\la , S_\De ) \ = \ S_\De^\infty \eeq
exists, we say that $Y$ is an $\om$-asymptotic symmetry for (the
submanifold $S_\De$ and hence) the equation $\De = 0$; and
similarly considering the limit for $\la \to -\infty$.

We have the following results, which we quote from \gcite{CGbook}
(Lemma IX.1 therein):

\medskip\noindent
{\bf Lemma \ref{sec:asymp}.1.} {\it Let $u = f_0 (x)$ be a solution
to the differential equation $\De_0$; and let the limits $f_{\pm
\infty}$, $\De_{\pm \infty}$ for the flows
$$ f_\la \ = \ \Phi (\la , f_0 ) \ , \ \ \De_\la \ = \
\Phi^{(n)} (\la, \De_0 ) $$ exist. Then $f_{\pm \infty}$ is a
solution to $\De_{\pm \infty}$.}

\medskip\noindent
{\bf Corollary.} {\it If $\De_0 \to \De_*$ for $\la \to \pm
\infty$, then under the flow $\Phi(\la,f)$ all solutions $f_0$ to
$\De_0$ go into solutions (not necessarily invariant) to $\De_*$
for $\la \to \pm \infty$.}

\medskip\noindent
{\bf Remark \ref{sec:asymp}.2.} We stress that here ``asymptotic''
refers to the flow parameter $\la$, i.e., to the flow $\Phi$ and
$\Phi^{(n)}$ associated to $Y$. This may or may not correspond to
asymptotic properties in time and/or space depending on the form
of the vector field $Y$ we are considering. \EOR

\subsection{Examples}

\medskip\noindent
{\bf Example \ref{sec:asymp}.1.} Consider the DS in $\R^2$ \beq
\cases{ \xd \ = \ x \ - \ A \, y \ - \ (x^2 + y^2 ) \, x & , \cr
\yd \ = \ y \ + \ A \, x \ - \ (x^2 + y^2 ) \, y & ; \cr} \eeq
this is symmetric under both time translations and spatial
rotations. Thus if we consider the two-parameters family of vector
fields \beq X \ = \ \a \,\pa_t \ + \ \b \ \( x \, \pa_y \, - \, y
\, \pa_x \) \ , \eeq all of these are symmetries for our DS. Note that for $\a > 0$ the action of $X$ gives translations \emph{forward} in time, while for $\a < 0$ we have translations \emph{backward} in time; similarly, the sign of $\b$ controls the direction of rotation in the $(x,y)$ plane. We assume $\a > 0$ for definiteness and ease of discussion.

The action of $X$ on the space $\Ga = \{ x_\la (t) , y_\la (t) \}$
of sections of $( R^2 \times R , \pi_0 , R )$ -- which we denote
by $\wt{X}$ -- is described by
\begin{eqnarray*}
\frac{d x_\la }{d \la} &=& - \, \a \ \[ 1 \ - \ (x^2 +y^2) \] \, x_\la \ - \ \( \b \ - \ \a \, A \) \, y_\la \ , \\
\frac{d y_\la }{d \la} &=& - \, \a \ \[ 1 \ - \ (x^2 +y^2) \] \, y_\la \ + \ \( \b \ - \ \a \, A \) \, x_\la \ . \end{eqnarray*}
Thus there is no fixed point for the action of $\wt{X}$ unless we
choose $\b = \a A$, i.e., $$ X \ = \ X_0 \ := \  \a \[ \pa_t \ + \
A \ \( x \, \pa_y \, - \, y \, \pa_x \) \] \ . $$ With this choice
instead, we get
\begin{eqnarray*}
\frac{d x_\la }{d \la} &=& - \, \a \ \[ 1 \ - \ (x^2 +y^2) \] \, x_\la \ , \\
\frac{d y_\la }{d \la} &=& - \, \a \ \[ 1 \ - \ (x^2 +y^2) \] \,
y_\la \ ;  \end{eqnarray*} thus \emph{all} solutions admit the vector field $X_0$ as an $\a$-asymptotic symmetry, and the vector field $X_0^* = -X_0$ as an
$\om$-asymptotic one (their role is reversed for $\a < 0$ in the definition of $X_0$). Note that the solutions with $x_0^2 +y_0^2$
equal to either zero or one also admit $X_0$ as a full symmetry.
\EOE

\medskip\noindent
{\bf Example \ref{sec:asymp}.2.} Consider the FKPP
(Fisher-Kolmogorov-Petrovskii-Piskunov) equation \gcite{Murray1,Murray2}
\beql{eq:FKPP} u_t \ = \ A \ u_{xx} \ + \ \eps \, u \, (1 - u )
\eeq for $u = u(x,t)$, with $A$ and $\eps$ positive real
constants; in biological applications, it is required that $u(x,t)
\ge 0$ for all $x$ and $t$. This obviously has some stationary
homogeneous (i.e., $u_x = 0$) states $u(x,t)= 0$ and $u(x,t) = 1$, with the former
being unstable and the latter stable against small perturbations.
It is known that if the initial datum is suitably concentrated, in
particular if it decays exponentially fast for $|x| \to \infty$, then
for $t \to \infty$ and $x \to \infty$ the solution is in the form
of a \emph{traveling front}
$$ u(x,t) \ \simeq \ \exp \[ - \, \frac{x \ - \ v \, t }{\la} \] \ ;
\ \ \ \la = \sqrt{A/\eps} \ , \ \ v \ = \ \sqrt{4 A \eps} \ . $$
It is convenient to rescale coordinates via
$$ t \to  \wt{t} \ = \ \eps \, t \ , \ \ x \to \wt{x} \ = \
\sqrt{\eps/A} \ x \ . $$ Using these coordinates -- and omitting the tildes
for ease of writing -- the equation reads \beql{eq:FKPPstand} u_t \ = \ u_{xx} \
+ \ u \ (1 \, - \, u ) \ , \eeq and the asymptotic solution is
$$ u (x,t) \ \simeq \ f_0 (x,t) \ = \ A \ \exp \[ - \, (x \, - \, 2 \, t ) \] \ . $$

As already mentioned, the asymptotic above describes the behavior of the solution
for large $t$ and $x$, which also means for small $u$. But for
small $u$ the equation is approximated by its linearization around
$u=0$, i.e., simply by \beql{eq:FKPPlin} u_t \ = \ u_{xx} \ + \ u \
; \eeq the \emph{ansatz} $u(x,t) =w(z) = w(x -2 t)$ transforms
this into \beql{eq:FKPPquot} w'' \ + \ 2 \, w' \ + \ w \ = \ 0 \ ,
\eeq with solutions
$$ w(z) \ = \ c_+ \, e^{-z} \ + \ c_- \, e^z \ . $$
We denote by $\mathcal{W}$ the space of such solutions; the
constants $(c_+,c_-)$ provide coordinates in this space. The
requirement that $w(z) \to 0$ for $z \to \infty$ (i.e. for $x \to
\infty$ at any finite $t$) yields $c_- = 0$ and hence $w(x) = c_+
e^{- z}$, which is just the $f_0 (x,t)$ given above.

The symmetry algebra of the linearized Eq \eqref{eq:FKPPlin}
is generated by
$$ X_0 \ = \ u \, \pa_u \ , \ \ X_1 \ = \ \pa_x \ , \ \ X_2 \ = \ \pa_t \ . $$
We also consider
$$ X_\pm \ := \ X_1 \ \mp \ \frac12 \ X_2 \ ; $$
note that $X_+= \pa_z $, $X_- (z)= 0$. Moreover, $[ X_0 , X_\pm]=0$.

The quotient (i.e., symmetry-reduced) Eq \eqref{eq:FKPPquot}
admits $X_0$ (now written as $X_0 = w \pa_w$) as a scaling symmetry, and is also invariant under
$X_+$. It is easily checked that the propagating front solutions
correspond to a subspace of $\mathcal{W}$  invariant under the
action of the group generated by the vector fields $X_0$ and
$X_+$.

As for asymptotic symmetries, the situation is less simple; one
can show the following, which we quote from \gcite{GMan2}: if we
consider a scaling vector field $X$ such that the limit for
$\la\to \infty$ of $\exp [ \la X]$ extracts the behavior for large
$x$ and $t$, and $\De_0$ is the FKPP equation, we set $\De_\la:=
e^{\la X} \De_0$. Then
$$ \lim_{\la\to \infty} \De_\la \ = \ \De_* $$
is the heat equation $u_t = u_{xx}$.

In fact, consider the most general scaling vector field,
$$ X_s \ = \  a \, x \, \pa_x \ + \ b \, t \, \pa_t \ + \ c \, u \, \pa_u \ ; $$ one of the constants $a,b,c$ can always be set to unity (provided it is nonzero), which amounts to a redefinition of the group parameter. Then choosing $\De_0$ the FKPP equation in its standardized form \eqref{eq:FKPPstand} we get for $\De_\la = \exp[X_s \la] \De_0$ the expression
$$ \De_\la \ = \ \la^{c-b} \ \[ u_t \ - \ \la^{b - 2 a} \, u_{xx} \ - \ \la^b \, u \ + \ \la^{b+c} \, u^2 \] \ . $$
In order to have a meaningful (and nonzero) limit for $\la \to \infty$ we set $b = c$, and we will choose $b=c < 0$; in order to have a nontrivial limit for $\la \to \infty$ we also set $a = b/2$. Moreover, we can reparametrize the flow setting $|b|=1$, i.e., $b = -1$. This yields
$$ \De_\la \ = \ u_t \ - \ u_{xx} \ - \ (u/\la)  \ + \ (u^2 / \la^2 )  \ , $$
which reduces to the heat equation for $\la \to \infty$. \EOE


\section{Symmetry of stochastic differential equations}
\label{sec:symmsto}

We have so far discussed symmetries of deterministic differential equations.
The symmetry approach has also been applied to \emph{stochastic}
differential equations \gcite{\stochsymmref}.


Here we will be specially concerned with
equations (as in the case of deterministic equation, here an equation might be of vector type, i.e., a system of scalar equations) of Ito type, i.e., in the form \beql{eq:Ito} d x^i \ = \
f^i (x,t) \, d t \ + \ \s^i_{\ k} (x,t) \, d w^k \ , \eeq
where$f^i$ and $\s^i_{\ k}$ are smooth functions. We will also
consider Stratonovich type equations \beql{eq:Strat} d x^i \ = \
b^i (x,t) \, d t \ + \ \s^i_{\ k} (x,t) \circ d w^k \ . \eeq

As it is well known, to each Ito equation there is an associated Stratonovich
equation (and conversely) which carries the same statistical
information (the correspondence between the two presents
some subtleties, see e.g., \gcite{Ikeda,Stroock} for a discussion of
these), in particular Eqs \eqref{eq:Ito} and \eqref{eq:Strat} are
in such a relation if their drifts satisfy \beql{eq:ItoS} f^i
(x,t) \ = \ b^i (x,t) \ + \ \frac12 \ \( \pa_k \s^{ij} \) \,
\s^k_{\ j} \ . \eeq

We will consider in particular the approach which parallels the
usual treatment of deterministic equations in the stochastic case
(to which we gave several contributions in recent years); for a
discussion of other approaches, including earlier attempts, see
e.g., the review paper \gcite{GGPR}. See also \gcite{DVMUG,DVMUGb} for an approach relating symmetries to Girsanov theory, and more generally \gcite{DVMU2,DVMU3,DVMU4,DVMU5,DVMTU} for a related concurrent approach.

While in the case of deterministic differential equations there exists a well developed \emph{geometric} theory, which allows to set the symmetry theory on firm geometrical basis, this does not hold for stochastic equations. Actually, the main characteristic of Ito differentials is that under diffeomorphisms they do \emph{not} transform by the familiar chain rule, but according to Ito rule. This shows that the lack of a geometrical setting for symmetries of Ito equations is not due to insufficient development of the theory, but to its intrinsic features. The situation is slightly different for Stratonovich equations. In this case the chain rule is preserved, so that a geometrical setting would be possible; but, as it is well known, the Stratonovich formulation presents several delicate points from the point of view of stochastic processes (many of them being actually associated to its more relevant advantage from the point of view of a physicist, i.e., the preservation of the time inversion symmetry), so that a formulation of the basic theory in terms of the Ito theory is preferable. The relation between symmetry theory for Ito and Stratonovich equations is thoroughly discussed e.g., in \gcite{Koz18b}; see also \gcite{Koz12} for earlier results.

\subsection{Symmetry of Ito equations}
\label{sec:symmIto}

Let us consider an Ito Eq \eqref{eq:Ito}; this involves the time variable $t$, the driving Wiener processes $w^k (t)$, and the spatial variables $x^i$, which evolve through a stochastic process depending on the realization of the $w^k (t)$ and described by the Ito equation itself.

Thus our problem lives in the space $\mathcal{E}$ of the variables $\{ t ; x ; w \}$ (note that $x$ and $w$ are vector variables); a blind application of the Lie approach would consider general vector fields in this space, i.e., vector fields of the form
\beq X \ = \ \tau (x,t,w) \, \frac{\pa}{\pa t} \ + \ \vphi^i (x,t,w) \, \frac{\pa}{\pa x^i} \ + \ h^k (x,t,w) \, \frac{\pa}{\pa w^k} \ . \eeq
It is clear that the action of such a vector field mixes all the variables, discarding the difference between driving and driven stochastic processes, and, even more, introducing a stochastic time. This kind of transformation is well known and widely used by probabilists, and also in the study of stochastic differential equations (including e.g., in the study of \emph{normal forms} for SDEs, which has close relations with symmetry considerations \gcite{ArnImk}); but here we are interested in transformations which leave the equation \emph{invariant}. Transformations which even change the nature of the variables are surely not leaving the equation invariant, and hence will not be considered.

If we want to leave the equation invariant, a necessary (but by no means sufficient) condition is that the nature of the involved variables is not changed. We should also ask the transformed $w^k$ variables to be still independent Wiener processes.
These requirements (see \gcite{GW18} for how the latter one restricts the form of the $h^k$ functions) lead to consider vector fields in $\mathcal{E}$ of the form
\beql{eq:Xa} X \ = \ \tau (t) \, \frac{\pa}{\pa t} \ + \ \vphi^i (x,t,w) \, \frac{\pa}{\pa x^i} \ + \ \( R^k_{\ \ell} w^\ell \) \, \frac{\pa}{\pa w^k} \ , \eeq
with $R$ a matrix in the Lie algebra of the conformal linear group $\mathrm{CL (n)}$.
These vector fields  are called \emph{admissible} in $\mathcal{E}$, and in the following we only consider such fields.

\medskip\noindent
{\bf Remark \ref{sec:symmsto}.1.} We recall $\mathrm{CL (n)} \approx {\bf R}_+ \times \mathrm{O(n)}$, where $\mathrm{O(n)}$ is the ($n$-dimensional) orthogonal group, and ${\bf R}_+$ represents here the dilation group. An alternative characterization of matrices $A \in \mathrm{CL (n)}$ is simply that they satisfy $A^+ A = \la^2 I$ for some $\la \in {\bf R}$. We are interested in matrices $R$ in the Lie algebra of $\mathrm{CL (n)}$; in practice, such an $R$ is the sum of a diagonal matrix which is a multiple of the identity (corresponding to the linear space spanned by the generator of the dilation group), and a skew-symmetric one (these correspond to the linear space spanned by the generators of the orthogonal part of the group). Thus $R$ has $1+n (n-1)/2$  independent elements.  \EOR

It is then straightforward to consider the change of variables induced by an infinitesimal action of $X$, i.e.,
$$ t \to t + \eps \tau \ , \ \ x^i \to x^i + \eps \vphi^i \ , \ \ w^k \to w^k + \eps R^k_{\ \ell} w^\ell \ , $$ and hence on the Eq \eqref{eq:Ito}. Our restrictions on the form of the vector field guarantees that the transformed equation is still an Ito one, which can hence be written as (recall the variables $t,x,w$ are now the transformed ones)
\beql{eq:Itotransf} d x^i \ = \
\^f^i (x,t) \, d t \ + \ \^\s^i_{\ k} (x,t) \, d w^k \ , \eeq
where the functions $\^f^i$ and $\^\s^i_{\ k}$ -- which can be readily computed in explicit terms -- are in general different from the original ones, $f^i$ and $\s^i_{\ k}$.

Requiring that the equation is left invariant under the action of $X$ means simply requiring that $\^f^i (x,t) = f^i (x,t)$ and that $\^\s^i_{\ k} (x,t) = \s^i_{\ k} (x,t)$. We have thus $n + n^2$ equations involving the unknown functions $\tau, \vphi^i, \s^i_{\ k}$,  which determine the vector fields which leave the Eq \eqref{eq:Ito} invariant, i.e., identify its (Lie-point) symmetries; these are thus called the \emph{determining equations}.

An important point is that the form \eqref{eq:Xa} for symmetry vector fields is actually still too general: albeit all vector fields of this form are admissible, not all of them are \emph{useful} when it comes to applying symmetry theory. In particular, as in the deterministic case, the main purpose of computing symmetries is to then apply them for reducing -- or even solving -- the stochastic equations. But to this purpose, only so called \emph{simple} symmetries are useful under the present theory \gcite{Koz1,Koz2,Koz3,KozB}. Simple symmetries are those not acting (but generally depending) on the time variable; i.e., those, in the notation of \eqref{eq:Xa}, with $\tau = 0$). Thus admissible simple symmetries have generators of the form
\beql{eq:Xas} X \ = \ \vphi^i (x,t,w) \, \frac{\pa}{\pa x^i} \ + \ \( R^k_{\ \ell} w^\ell \) \, \frac{\pa}{\pa w^k} \ , \eeq
with $R$ as discussed above.

We stress that this restriction to simple (admissible) symmetries will be given for granted in the following; it will also simplify a number of matters.

Some further nomenclature within simple symmetries may also be useful: when $R=0$ and the $\vphi^i$ do \emph{not} depend on the $w$ variables, we have \emph{deterministic} symmetries; when $R=0$ and (at least some of) the $\vphi^i$ \emph{do} actually depend on the $w$ variables, we have \emph{random} symmetries; and finally when $R\not= 0$ (and hence the $X$ has a component along the $w$ variables) we have \emph{W-symmetries}.

\medskip\noindent
{\bf Remark \ref{sec:symmsto}.2.} There is a substantial difference in computing the action of a vector field $X$ on a stochastic equation with respect to the case of a standard (i.e., deterministic) ordinary differential equation. In the latter case the ODE defines a vector field $Z$ in a suitable jet bundle $J$, and we compute how the vector field $X$ and its prolongations act on the vector field $Z$; in the case of stochastic Ito equations, we have to compute how $X$ acts on the functions $f^i (x,t)$, $\s^i_{\ k} (x,t)$ and on the differentials $d x^i$, $d t$, $d w^k$. Thus, roughly speaking, we are dealing with one-forms rather than with vector fields. This might be compared with recent work by Anco and Wang \gcite{Anco} taking the same point of view for deterministic ODEs. \EOR

When we write down the explicit expressions for $\^f^i$ and $\^\s^i_{\ k}$, we also obtain the explicit expressions for the determining equations. We set this in the form of a Proposition; this is quoted from \gcite{GW18}, and we refer to there for a proof.

Here and below\ the symbol $\Delta$ denotes the \emph{Ito Laplacian},
\beql{eq:ItoLap} \Delta F \ := \ \sum_{k,\ell} \de^{k \ell} \frac{\pa^2 F}{\pa w^k \pa w^\ell} \ + \ \sum_{i,j} ( \s \s^T )^{ij} \, \frac{\pa^2 F}{\pa x^i \pa x^j} \ + \ 2 \, \sum_{j,k} \s^{jk} \, \frac{\pa^2 F}{\pa x^j \pa w^k} \ . \eeq

\medskip\noindent
{\bf Proposition \ref{sec:symmsto}.1.} {\it The determining equations for simple  symmetries of the Ito Eq \eqref{eq:Ito} are
\begin{eqnarray}
\frac{\pa \vphi^i}{\pa t} \ + \ f^j \, \frac{\pa \vphi^i}{\pa x^j} \ - \ \vphi^j \, \frac{\pa f^i}{\pa x^j} \ + \ \frac12 \, \Delta \vphi^i &=& 0 \ , \label{eq:deteqIto1} \\
\frac{\pa \vphi^i}{\pa w^k} \ + \ \s^j_{\ k} \, \frac{\pa \vphi^i}{\pa x^j} \ - \ \vphi^j \, \frac{\pa \s^i_{\ k}}{\pa x^j} \ - \ \s^i_{\ m} \, R^m_{\ k} &=& 0 \ . \label{eq:deteqIto2} \end{eqnarray}}

Obviously, several special cases are obtained by considering special forms of the functions $\tau$, $\vphi^i$ and $R$. We will not enter into this discussion, and just refer the reader to \gcite{GGPR,GW18}.

We will however stress that solutions with $\tau = 0$ (i.e., such that the time variable is not acted upon by the vector field $X$) are known as \emph{simple} symmetries, and are specially useful in that the symmetry reduction based on these is specially simple.

\medskip\noindent
{\bf Remark \ref{sec:symmsto}.3.} The determining Eqs \eqref{eq:deteqIto1}, \eqref{eq:deteqIto2} are a set of $n+n^2$ equations for the $n$ functions $\{ \vphi^1 , ... , \vphi^n \}$ and the $n \times n$ matrix $R$, having (as noted above) $n (n-1)/2 +1$ independent elements. In order to concretely use the information about the symmetry, it does not suffice to prove that a solution exists, but we need of course to determine it in an explicit way. \EOR

\subsection{Ito versus Stratonovich equations}
\label{sec:ItoStrat}

As it is well known -- and recalled above -- the Ito Eq \eqref{eq:Ito} is associated with the Stratonovich Eq \eqref{eq:Strat}, the coefficients of the two being related by \eqref{eq:ItoS}. With this condition, \eqref{eq:Strat} carries the same statistical information as \eqref{eq:Ito}; see e.g., \gcite{Stroock} for more detail. From the point of view of symmetry, there is a substantial difference: while the Ito equations transforms, under the action of a diffeomorphism and thus in particular under the flow generated by a vector field, as described by the \emph{Ito rule}, a Stratonovich equation transforms under the usual chain rule and thus has a geometrical meaning.

It is quite natural to wonder if symmetries of an Ito equation and of the associated Stratonovich equation are related or coincide. The question was first tackled by Unal \gcite{Unal}, and then fully solved by Kozlov \gcite{Koz18b}, who showed that deterministic and random symmetries of an Ito and of the associated Stratonovich stochastic equations are just the same -- as indeed natural given the equivalence of the two formulations.

The situation is slightly different for W-symmetries; the relations of these for an Ito and the associated Stratonovich equation are discussed in \gcite{GW18}, see Section VII (and in particular Section VII.C) therein.

One could of course also analyze symmetries of a Stratonovich Eq \eqref{eq:Strat} along the lines of Section \ref{sec:symmIto}, with the same limitation on the form if the admitted vector fields; in this way one would obtain the determining equations for (Lie-point) symmetries of a Stratonovich equation. See again \gcite{GW18} for a proof to the following statement.

\medskip\noindent
{\bf Proposition \ref{sec:symmsto}.2.} {\it The determining equations for simple symmetries of the Stra\-to\-no\-vich Eq \eqref{eq:Strat} are
\begin{eqnarray}
\frac{\pa \vphi^i}{\pa t} \ + \ b^j \, \frac{\pa \vphi^i}{\pa x^j} \ - \ \vphi^j \, \frac{\pa b^i}{\pa x^j} &=& 0  \ , \label{eq:deteqStrat1} \\
\frac{\pa \vphi^i}{\pa w^k} \ + \ \s^j_{\ k} \, \frac{\pa \vphi^i}{\pa x^j} \ - \ \vphi^j \, \frac{\pa \s^i_{\ k}}{\pa x^j} & = & \s^i_{\ m} \, R^m_{\ k} \ .  \label{eq:deteqStrat2} \end{eqnarray}}
Note that \eqref{eq:deteqStrat2} coincides with \eqref{eq:deteqIto2}; as for \eqref{eq:deteqStrat1}, this coincides with \eqref{eq:deteqIto1} when we impose the correspondence \eqref{eq:ItoS}. See also \gcite{GL1,GL2,Koz18a,Koz18b} for further detail.

\subsection{Adapted variables and integration of stochastic equations}
\label{sec:adapt}

In the view of Sophus Lie, the symmetry analysis of differential equations was not a mathematical curiosity, but a concrete method to obtain order/dimensional reduction, and when possible exact solutions, of the differential equations.

The same applies for symmetry analysis of stochastic equations: knowledge of their symmetries allows to obtain a reduction, and when we have enough symmetries (e.g., one symmetry for a scalar equation) there is a concrete method to obtain a complete solution of the equation.

\medskip\noindent
{\bf Remark \ref{sec:symmsto}.4.} It is maybe worth specifying what we mean by this, i.e., by a complete solution of an Ito equation. We mean an explicit correspondence between a concrete realization of the driving ($n$-dimensional) Wiener process and the corresponding realization of the ($n$-dimensional) stochastic process $x(t)$. Example of this correspondence are provided below in this subsection. \EOR

As for deterministic equations, the key to integration of a symmetric equation lies in passing to symmetry-adapted variables (an alternative method uses \emph{invariants}, and will be discussed in the next Section). In these, the equation takes a specially simple form and is integrated in an elementary way; that is, the difficulty of integrating the equation is moved to the difficulty of determining the symmetry and the symmetry-adapted variables.

In fact, once we have determined (simple) symmetries of a given Ito equation, these can be used pretty much in the same way as in the case of deterministic equations to integrate or reduce the equation at hand.

\medskip\noindent
{\bf Proposition \ref{sec:symmsto}.3.} {\it Let the Eq \eqref{eq:Ito} for $x \in M \approx {\bf R}^n$ admit a $r$-dimensional solvable algebra $\mathcal{G}$ of simple \emph{deterministic} Lie symmetries. Then, passing to $\mathcal{G}$-adapted variables, the equation can be reduced to a stochastic equation for $y \in M/\mathcal{G} \approx {\bf R}^{n-r}$ plus a set of $r$ reconstruction equations; the latter amount to simple Ito integrations. In the case $r=n$, the equation is explicitly integrated.}

\medskip\noindent
{\em Proof.} This is given and proved e.g., in \gcite{GL2}; it is however worth sketching the proof. The first step, i.e., the determination of (simple) Lie symmetries, has already been mapped in the previous part of this section to the problem of solving the determining equations. Assume that a symmetry vector field $X$ -- written in the form \eqref{eq:Xas} -- has been found. We want then to pass from $(x,t;w)$ to new variables $(y,\theta;z)$, where the $z^k$ are still independent Wiener processes, such that in the new variables $X$ reads simply as $X =\pa / \pa y^n$. In fact, if this is the case the Ito equation will be written, in the new variables, as
\beq d y^i \ = \ F^i (y,\theta) \, d \theta \ + \ S^i_{\ k} (y,\theta) \, d z^k
\eeq with both the $F^i$ and the $S^i_{\ k}$ being independent of $y^n$. The equation decouples then into an $(n-1)$-dimensional Ito equation plus a ``reconstruction equation'', which is written in terms of an Ito integral as
\beq y^n (t) \ = \ y^n (0) \ + \ \int_0^t F^n [ y(\theta) , \theta] \, d \theta \ + \ \int_0^t S^n_{\ k} [y (\theta),\theta] \, d z^k \ . \eeq
In the case of a scalar equation, this gives directly a full solution of the equation.

When we have a higher dimensional equations and several symmetries spanning a solvable Lie algebra, we can proceed step by step by reducing the dimension of the system provided we operate according to the structure of the Lie algebra, i.e., according to the derived series for $\mathcal{G}$, as in the case of deterministic equations \gcite{\symmref}. \EOP

We stress that for simple Lie-point symmetries, i.e., vector fields of the form
\beq X \ = \ \vphi^i (x,t) \ \pa_i \ , \eeq there is a simple way to relate the sought-for change of variables to the symmetry $X$ \gcite{Koz1,Koz2,Koz3}. First of all we note that in this case the time variable can be left unaffected by the change of variables. It suffices then to require that
$$ X \ = \ \vphi^i (x,t) \, \frac{\pa}{\pa x^i} \ = \ \vphi^i [ x(y),t] \, \frac{\pa y^\ell}{\pa x^i} \,\frac{\pa}{\pa y^\ell} \ = \ \frac{\pa}{\pa y^n} \ ;  $$
this requires to solve the linear equations
$$ \cases{\vphi^i (x,t) \, \( \pa y^\ell / \pa x^i \) \ = \ 0 & ($\ell = 1,...,n-1$) \cr
\vphi^i (x,t) \, \( \pa y^n / \pa x^i \) \ = \ 1 & , \cr} $$
which are solved by the classical method of characteristics.

In particular, for $n=1$ we have to solve only the last equation, and its solution is promptly written as
$$ y \ = \ \int \frac{1}{\vphi (x,t)} \ d x \ . $$

\medskip\noindent
{\bf Remark \ref{sec:symmsto}.5.} The reader can wonder why this Proposition considers only \emph{deterministic} symmetries. The reason is that in the case of random or  $W$-symmetries, one is \emph{not} guaranteed that the transformed equation is still of Ito type (this depends on a certain compatibility condition being satisfied, see \gcite{GL2,GW18}). Note that the problem is specially present when attempting multiple reduction. In fact, the first reduced equation could be not of Ito type, and in this case our theory can not say anything about successive reductions.

We also stress that in the scalar case the reduced equation, albeit not of Ito type, could be integrable; in this case its solution provides -- inverting the transformation -- the solution to the original equation.

Finally, we note that in \gcite{GW18}, see in particular Sect. V therein, condition ensuring the symmetry reduced equations are still of Ito type are discussed in some detail (in any case, one can perform the reduction and verify if the reduced equation is Ito or otherwise). When all the reduced equations are of Ito type, Proposition 3 extends also to random or W-symmetries. \EOR

\subsection{The Lie algebraic structure of symmetries for a stochastic equation}
\label{sec:LAS}

It is well known that symmetries of a deterministic differential equations form a Lie algebra \gcite{\symmref} (for an explicit proof, see e.g., Sect. 2.3 (and Sect. 5.1 for the case of generalized symmetries) in ref.\gcite{Olver1}). The same holds for stochastic equations, at least for deterministic or random symmetries; note this includes the class of symmetries (deterministic ones) which are guaranteed to allow multiple reductions (see Remark \ref{sec:symmsto}.5 above):

\medskip\noindent
{\bf Proposition \ref{sec:symmsto}.4.} {\it The set of simple deterministic or random symmetries of a given Ito equation has the standard Lie algebra structure, i.e., if $X,Y$ are symmetries of a given equation, their commutator $Z =[X,Y]$ is also a symmetry for the same equation.}

\medskip\noindent
{\em Proof.} This is Theorem 4.3 in \gcite{Koz18a}; see there for the proof. Alternatively, note that symmetries of an Ito equation correspond to symmetries of the associated Stratonovich equation; for the latter the standard chain rule holds, so it is obvious that symmetries have the standard Lie algebra structure, proceeding just as in the case of deterministic equations. \EOP

\medskip\noindent
{\bf Remark \ref{sec:symmsto}.6.} It should be noted that \gcite{GS17} contains a statement in contrast with Proposition \ref{sec:symmsto}.4 (see Sect. IX.A therein). That statement concerns transformations of a rather more general form than the one considered here\footnote{In fact, far too general transformations, in particular functions $h(x,t,w)$, are allowed at that point of the discussion in \gcite{GS17}; restricting to $h(x,t,w) = R w$, as we do here,  produces the right result.} and it is thus wrong if seen in the present context, while Proposition \ref{sec:symmsto}.4 above is correct. \EOR

\medskip\noindent
{\bf Remark \ref{sec:symmsto}.7.} Our Proposition \ref{sec:symmsto}.4 does not say anything concerning W-sym\-me\-tri\-es. In fact, the following Examples will show that in some cases these also form a Lie algebra (or possibly even a Lie module), while in other cases this is not the case. In other words, no general statement is possible about the Lie algebraic structure of W-symmetries of a given equation. \EOR

\subsection{Stochastic versus diffusion equations}
\label{sec:FP}

Finally, we recall that for any Ito Eq \eqref{eq:Ito} there is an associated diffusion (Fokker-Planck) equation
\beql{eq:FP} u_t \ = \ \ A^{ij} \, u_{ij} \ - \ B^i \, u_i \ - \ C \, u  \ , \eeq
describing the evolution of the density $u (x,t)$ in time; here $u_i := (\pa u / \pa x^i)$ and similarly $u_{ij} := (\pa^2 u / \pa x^i \pa x^j)$, and the coefficients in the Fokker-Planck Eq \eqref{eq:FP} are related to those in the Ito Eq \eqref{eq:Ito} by
\beq A^{ij} \ = \ \frac12 \, (\s \, \s^T)^{ij} \ , \ \ B^i \ = \ f^i \ - \ \pa_j ( \s \s^T)^{ij} \ , \ \ C \ = \ (\pa_i f^i) \ - \ \frac12 \, \pa^2_{ij} (\s \s^T)^{ij} \ . \eeq

\medskip\noindent
{\bf Remark \ref{sec:symmsto}.7.} The Fokker-Planck equation is often also written, more synthetically, as
\beql{eq:FPshort} u_t \ = \ \frac12 \, \frac{\pa^2}{\pa x^i \pa x^j} \[ \s \s^T \ u \] \ - \ \frac{\pa}{\pa x^i} \, \[ f^i \, u \]  . \eeq
Combining this equation with those for the transition from an Ito equation to the associated Stratonovich one, see Eq \eqref{eq:ItoS}, it is immediate to get the Fokker-Planck equation associated to a Stratonovich equation. \EOR

\medskip\noindent
{\bf Remark \ref{sec:symmsto}.8.} We also recall that, as discussed in \gcite{Koz12,GRQ1,GGPR}, any symmetry of the Ito Eq \eqref{eq:Ito} leads (adding a standard term proportional to $\pa_u$) to a symmetry of the associated Fokker-Planck Eq \eqref{eq:FP}, but the converse is not true. This is rather natural, given that different Ito equations can have the same associated Fokker-Planck equation; in fact, symmetries of a Fokker Planck equation will in general map these different Ito equations one into the other. We refer to \gcite{Koz12,GRQ1,GRQ2,GGPR} for further detail and examples. \EOR

\subsection{Examples}

We consider here some examples illustrating the results discussed in this Section. We recall that we are considering \emph{simple} admissible symmetries or W-symmetries; that is, vector fields of the form \eqref{eq:Xas} with $R$ a constant matrix of the form discussed in Sect. \ref{sec:symmIto}.

\medskip\noindent
{\bf Example \ref{sec:symmsto}.1.}
Consider the equation
$$ d x \ = \ t \ d w \ . $$ The determining equations are then
$$ \vphi_t \ + \ \frac12 \, \Delta (\vphi) \ = \ 0 \ , \ \
\vphi_w \ + \ t \, \vphi_x \ - \ R \, t \ = \ 0 \ ; $$ their general solution is
$$ \vphi \ = \ R \, x \ + \ k \ , $$ with $k$ a constant. Thus we have two symmetries (a deterministic one and a W-symmetry), i.e.,
$$ X \ = \ \pa_x \ , \ \ Y \ = \ R \, x \, \pa_x \ + \ R \, w \, \pa_w \ . $$
It is immediate to check that
$$ [X,Y] \ = \ R \, \pa_x $$ is still a symmetry: thus in this case the set of (simple) symmetries, involving both ``ordinary'' and W-symmetries, is indeed a Lie algebra. \EOE

\medskip\noindent
{\bf Example \ref{sec:symmsto}.2.} Consider the scalar Ito equation
\beql{eq:GBM} d x \ = \ \a \, x \, dt \ + \ \b \, x \, d w \ , \eeq
where $\a$ and $\b$ are real constants. This describes the  \emph{geometric Brownian motion} in one dimension, and it is well known that this equation can be integrated (see e.g., \gcite{Oksendal}, Chapter 5). Note that if we start with $x(0) > 0$, we are guaranteed to have $x(t) \ge 0$ for all times.  The determining equations \eqref{eq:deteqIto1}, \eqref{eq:deteqIto2} for simple symmetries\footnote{Recall this means $\tau = 0$ in terms of \eqref{eq:Xa}, i.e., that $X$ should be of the form \eqref{eq:Xas}.} of \eqref{eq:GBM} are
\begin{eqnarray}
\vphi_t \ + \ \a \, x \, \vphi_x \ - \ \a \, \vphi \ + \ \frac12 \, \Delta \vphi &=& 0 \ , \label{eq:GBM1} \\
\vphi_w \ + \ \b \, x \, \vphi_x \ - \ \b \, \vphi \ - \ \b \, x \, R &=& 0 \ . \label{eq:GBM2} \end{eqnarray}
Solving \eqref{eq:GBM2} yields
$$ \vphi (x,t;w) \ = \ x \ \[  R \, \log (x) \ + \ \psi (z,t) \]  \ ; \ \ \ \ z \ := \ w \ - \ \b^{-1} \ \log(x) \ . $$
Inserting this into \eqref{eq:GBM1}, the latter reads
$$ \( \frac{x}{2 \b} \) \ \[ \b \, (\b^2 \ + \ 2 \, \a  ) \, R \ + \ 2 \, \b \, \psi_t \ + \ ( \b^2 \ - \ 2 \, \a ) \, \psi_z \] \ = \ 0 \ ; $$
the solution to this equation is
$$ \psi (z,t) \ = \ \frac{ \b \, (2 \a \, + \, \b^2 ) \, R z \ + \ (2 \a \, - \, \b^2 ) \, Q (\zeta )}{(2 \a \, - \, \b^2 ) } \ , $$
having defined
$$ \zeta \ = \ \frac{(2 \, \a \ - \ \b^2 ) \, t \ + \ 2 \, \b \, w \ - \ 2 \, \log (x)}{2 \, \a \ - \ \b^2 } \ . $$
Going back to evaluate $\vphi$, we thus we have symmetries depending on the arbitrary function $Q$ and on the arbitrary constant $R$
\beql{eq:Xexa51} X \ = \ \[ x  \ \( \frac{\b}{2 \a - \b^2} \, \( (2 \a + \b^2 ) \, w \ - \ 2 \, \b \, \log (x) \) \ R \ + \ Q (\zeta ) \) \] \ \pa_x \ + \ \( R \, w \) \, \pa_w \ . \eeq
The simplest choice is of course $R=0$, $Q(\zeta) = 1$, which gives $\vphi = x$, i.e. the scaling symmetry
$$ X_0 \ = \ x \ \pa_x \ . $$

The associated integrating change of variables is simply
$$ y \ = \ \int \frac{1}{\vphi} \ d x \ = \ \int \frac{1}{x} \ d x \ = \ \log (x) \ ; $$
In fact, Ito formula for the change of variables gives immediately (using \eqref{eq:GBM} for the time evolution)
$$ d y \ = \ (\a \ - \ \b^2/2 ) \ dt \ + \ \b \ dw \ , $$ which yields at once (setting for ease of writing $t_0 = 0$ and $w(0)=0$)
$$ y(t) \ = \ y_0 \ + \ (\a - \b^2/2 ) \ t \ + \ \b \, w(t) \ , $$
and hence
$$ x(t) \ = \ \exp[y(t)] \ = \ x_0 \ \exp \[ (\a - \b^2/2 ) \ t \ + \ \b \, w(t) \] \ . $$

We can use this Example also to illustrate the Lie algebraic structure of the set of symmetries. Let us first set $R=0$, i.e., consider random symmetries, and two symmetries of the general form \eqref{eq:Xexa51} with $R=0$, say
$$ X \ = \ x \, P (\zeta) \ \pa_x \ , \ \ \ Y \ = \ x \, Q (\zeta ) \ \pa_x \ . $$
An explicit computation shows that
$$ Z \ := \ [X,Y] \ = \ x \ H(\zeta) \ \pa_x $$ where the function $H$ is given explicitly by
$$ H(\zeta) \ := \ \frac{1}{\a - \b^2/2} \ \[ P' (\zeta) \, Q(\zeta) \ - \ P(\zeta) \, Q' (\zeta) \] \ . $$ This is still of the functional form \eqref{eq:Xexa51} with $R=0$, and hence is immediately known to be still a (random) symmetry for our Eq \eqref{eq:GBM}.

On the other hand, let us consider the general case, i.e., let us \emph{not} enforce $R=0$. We set a simplified notation writing
$$ \mu (x,w) \ = \ \frac{\b}{2 \a - \b^2} \( (2 \a + \b^2) w - 2 \b \log (x) \) \ , $$ and consider vector fields
\begin{eqnarray*} X &=& x \ \( \mu (x,w) \, R_1 \ + \ P (\zeta) \) \, \pa_x \ + \ (R_1 w) \, \pa_w \ , \\
Y &=& x \ \( \mu (x,w) \, R_2 \ + \ Q (\zeta) \) \, \pa_x \ + \ (R_2 w) \, \pa_w \ ; \end{eqnarray*}
by our discussion above, they are symmetries of \eqref{eq:GBM}; we could actually also take $R_1 = R_2 = R \not= 0$.
Then an explicit computation yields
$$ Z \ = \ [X,Y] \ = \ x \ h (x,w;\zeta) \ \pa_x $$
where the function $h$ is given by
\begin{eqnarray*} h(x,w;\zeta) &:=& \frac{1}{(\a - \b^2/2)^2} \ \[ \b^2 \ \( \b w - \log (x) \) \ \( R_2 \, P' (\zeta) \ - \ R_1 \, Q' (\zeta) \) \right. \\
& & \left. \ + \ \( \a - \b^2/2\) \[ (R_1 Q - R_2 P ) \b^2 \ + \ P' (\zeta) \, Q(\zeta) \ - \ P (\zeta) \, Q' (\zeta ) \] \] \ . \end{eqnarray*}
The vector field $Z$ thus defined is \emph{not} of the form \eqref{eq:Xexa51} and hence is \emph{not} a symmetry of the Eq \eqref{eq:GBM}.

This shows that W-symmetries of a given equation, contrary to simple deterministic and random symmetries, in general do \emph{not} form a Lie algebra. \EOE

\medskip\noindent
{\bf Example \ref{sec:symmsto}.3.} Consider the \emph{stochastic logistic equation with environmental noise}
\beql{eq:SLE} d x \ = \ \( \a \, x \ - \ \b \, x^2 \) \ d t \ + \ \ga \, x \, d w \ . \eeq
Here $\a,\b,\ga$ are nonzero real constants; in the biological applications they have  positive sign, but for our present purposes this is not relevant.

We will again look at the case with $R=0$, i.e., search for symmetries of the form $X = \vphi (x,t;w) \pa_x$. In this case the second determining equation \eqref{eq:deteqIto2} reads
$$ \vphi_w \ + \ \ga \ \( x \, \vphi_x \ - \ \vphi \) \ = \ 0 \ , $$ which yields immediately
$$ \vphi (x,t,w) \ = \ x \ \psi (z,t) \ ; \ \ \ z \ = \ w \ - \ \ga^{-1} \ \log (x) \ . $$ Then the first determining Eq \eqref{eq:deteqIto1} splits into two equations (for the coefficients of $x$ and $x^2$ in the resulting expression):
\begin{eqnarray*}
\psi_t &=& \( \frac{\a}{\ga} \ - \ \frac{\ga}{2} \) \ \psi_z \\
\psi_z &=& - \, \ga \ \psi \ . \end{eqnarray*}
These are easily solved, and we get in the end
$$ \vphi (x,t,w) \ = \ x^2 \ \exp \[ - \, \( A \, t \ + \ \ga \, w \) \] \ , \ \ \ A \ := \ \a \ - \ \ga^2 /2 \ . $$
Note this yields a random symmetry
$$ X \ = \ x^2 \ \exp \[ - \, \( A \, t \ + \ \ga \, w \) \] \ \pa_x \ .  $$

The associated change of variables and its inverse are given by
 \begin{eqnarray} y &=& \int \frac{1}{\vphi (x,t,w)} \ d x \ = \  - \, \frac{1}{x} \ \exp[ A \, t \ + \ \ga \, w ] \ ; \label{eq:SLEcvd} \\
 x &=& - \, \frac{1}{y} \ \exp [ A \, t \ + \ \ga \, w ] \ . \label{eq:SLEcvi} \end{eqnarray}
When we express the evolution of $y$ using Ito's rule and \eqref{eq:SLE}, we get $$ d y \ = \ - \, \b \ \exp \[ A \, t \ + \ \ga \, w \] \ d t \ ; $$
note this is \emph{not} an Ito equation. It can however be directly integrated to give
$$ y(t) \ = \ y_0 \ - \ \b \ \int_0^t \exp \[ A \tau + \ga w (\tau) \] \ d \tau \ . $$ Transforming back to the original variable $x$ via \eqref{eq:SLEcvi} we obtain the solution to \eqref{eq:SLE} in closed -- albeit involved -- form. The reader can see \gcite{GSLE} for numerical experiments confirming the analytical procedure. \EOE

\medskip\noindent
{\bf Example \ref{sec:symmsto}.4.} Consider the equation
\beql{eq:NE3} d x \ = \ x \, e^{- t} \, d t \ + \ x \ d w \ . \eeq
(note that $x(0)>0$ entails $x(t) \ge 0$ for all $t \ge 0$.)
This admits as symmetry
\beql{eq:Xexa53} X \ = \ x \ F (\zeta )  \ \pa_x \ , \eeq
where $F$ is an arbitrary function of
$$ \zeta \ := \ e^{- t } \ + \ \frac{t}{2} \ - \ w \ + \ \log (x) \ . $$
The simplest choice is of course $F (\zeta) = 1$, i.e., $$ X \ = \ x \ \pa_x \ ; $$ note this is a \emph{deterministic} symmetry. The associated change of variables is just $y = \log (x)$, and with this our Eq \eqref{eq:NE3} is mapped into
$$ d y \ = \ \( e^{-t} \, - \, 1/2 \) \ dt \ + \ dw \ , $$ which is directly integrated to
$$ y(t) \ = \ y (t_0) \ - \ \[ \frac{(t- t_0)}{2} \ + \ \( e^{-t} - e^{- t_0} \) \] \ + \ \( w(t) \, - \, w(t_0 ) \) \ . $$
This yields the solution to our Eq \eqref{eq:NE3} upon inversion of the change of variables, i.e., setting $x = e^y$.

Suppose we do not want to make this simplest choice, and go for the next simpler one; this will produce a \emph{random} symmetry. So let us choose $F(\zeta) = \zeta$, and hence $$ X \ = \ x \ \[ e^{- t } \ + \ \frac{t}{2} \ - \ w \ + \ \log (x) \] \ \pa_x \ . $$  Then the associated change of variables is
$$ y \ = \ \log \[ e^{-t} \, + \, \frac{t}{2} \, - \, w \, + \, \log (x)  \] \ + \ \b (t,w) \ . $$ the equation for $y$ is then
$$ d y \ = \ \( \b_t \ + \ \frac12 \, \b_{ww} \) \, d t \ + \ \b_w \, d w \ . $$ Thus we should choose $\b (t,w) = b (t) + c w$ in order to get an Ito equation. With this choice, the initial equation is mapped into
$$ d y \ = \ \[ b' (t) \] \ d t \ + \ c \ d w \ , $$
which is readily integrated to yield
$$ y(t) \ = \ y(t_0) \ + \ \[ b(t) \ - \ b(t_0) \] \ + \ c \[ w(t) \ - \ w(t_0)\] \ . $$
Inverting the change of coordinates -- which is now a more complex operation than in the previous case -- we obtain the solution for $x(t)$.

We can again check that the symmetries, i.e., vector fields of the form \eqref{eq:Xexa53} (which are random symmetries), form a Lie algebra. In this case, it is in fact immediate to check that if
$$ X \ = \ x \ F(\zeta) \ \pa_x \ , \ \ \ Y \ = \ x \ G(\zeta) \ \pa_x \ , $$
then we have
$$ Z \ := \ [X,Y] \ = \ x \ H (\zeta) \ \pa_x $$
where we have defined
$$ H(\zeta) \ = \ F' (\zeta) \ G(\zeta ) \ - \ F(\zeta) \ G(\zeta) \ . $$ Thus $Z$ is of the form \eqref{eq:Xexa53}, and hence a symmetry for Eq \eqref{eq:NE3}, as claimed.  \EOE

\medskip\noindent
{\bf Example \ref{sec:symmsto}.5.} Consider the two-dimensional equation (here and elsewhere in concrete examples, we write vector indices as lower ones, to avoid any possible confusion with squares)
\begin{eqnarray}
d x_1 &=& \[ e^{x_1} \ - \ \frac12 \, e^{- 2 x_1} \] \, d t \ + \ e^{- x_1 } \, d w_1 \nonumber \\
d x_2 &=& \frac12 \, e^{x_2} \, \[ 2 \, e^{x_1}  \ + \ e^{x_2} \ + \ e^{2 x_1 + x_2} \] \, d t \ - \ e^{x_1+x_2} \, d w_1 \ - \ e^{x_2} \, d w_2 \ . \label{eq:GL2dim} \end{eqnarray}
It is easily checked that this admits the symmetry vector field
\beq X \ = \ - \, e^{x_2} \, \pa_2 \ ; \eeq symmetry-adapted variables are
\beql{eq:GL2dimcv} y_1 \ = \ \exp [ x_1 ] \ , \ \ y_2 \ = \ \exp [ - x_2 ] \ ; \eeq
in these variables, the vector field is
$$ X \ = \ \pa / \pa y_2  $$ and the eq. \eqref{eq:GL2dim} reads simply
\begin{eqnarray}
d y_1 &=& y_1^2 \, dt \ + \ d w_1 \nonumber \\
d y_2 &=& - y_1 \, dt \ + \ y_1 \, dw_1 \ + \ dw_2 \ . \end{eqnarray}
That is, we have an \emph{autonomous} Ito equation for $y_1$; \emph{if} this is solved, say with $y_1 = \Phi [ t, w_1 (t) ]$, \emph{then} the equation for $y_2$ is immediately integrated to give
$$ y_2 (t) \ = \ y_2 (0) \ - \ \int_0^t \Phi [t,w_1 (t)] \, dt \ + \ \int_0^t \Phi [ t , w_1 (t) ] \, d w_1 (t) \ + \ \int_0^t d w_2 (t) \ . $$
The solution to the initial problem \eqref{eq:GL2dim} is then obtained by inverting the change of variables \eqref{eq:GL2dimcv}. \EOE

\medskip\noindent
{\bf Example \ref{sec:symmsto}.6.}
Consider, as in \gcite{GW18}, the two-dimensional isotropic linear stochastic equation \beql{eq:2DISO} d x^i \ = \ \la \, x^i \, d t \ + \ \mu \, d w^i \ \ \ \ (i=1,2) \ . \eeq
This was considered, for what concerns W-symmetries in \gcite{GW18} (this was  Example 7 therein), but only W-symmetries were considered in that work; we will thus have to derive the full symmetry algebra.

We write, for ease of notation, $R$ in the form
$$ R \ = \ \left( \matrix{ r_0 & r_1 \cr - r_1 & r_0 \cr} \right) \ . $$
Solving the $\s$-determining equations for $\vphi^i (\xb, t ; \wb)$ yields
\begin{eqnarray*} \vphi^1 (x_1,x_2,t;w_1,w_2) &=& r_0 \, x_1 \ + \ r_1 \, x_2 \ + \ \psi^1 (z_1,z_2,t) \ , \\
\vphi^2 (x_1,x_2,t;w_1,w_2) &=& - r_1 \, x_1 \ + \ r_0 \, x_2 \ + \ \psi^2 (z_1,z_2,t) \ ; \\
z_i &:=& w_i \ - \ x_i / \mu \ . \end{eqnarray*}
With this, the $f$-determining equations read
\begin{eqnarray*}
\frac{\pa \psi^1}{\pa t} & = & \frac{\la}{\mu} \, \( x_1 \, \frac{\pa \psi^1}{\pa z_1} \ + \ x_2 \, \frac{\pa \psi^1}{\pa z_2} \) \ + \ \la  \, \psi^1 \ , \\
\frac{\pa \psi^2}{\pa t} & = & \frac{\la}{\mu} \, \( x_1 \, \frac{\pa \psi^2}{\pa z_1} \ + \ x_2 \, \frac{\pa \psi^2}{\pa z_2} \) \ + \ \la  \, \psi^2 \ . \end{eqnarray*}
Note that the $\psi$ are functions of the $z_i$ variables alone; thus coefficients of the $x_i$ must vanish separately, and we get
$$ \psi^i (z_1,z_2,t) \ = \ e^{\la t} \ c_i $$ where $c_i$ are arbitrary constants.

In conclusion, the symmetry algebra $\mathcal{G}$ for \eqref{eq:2DISO} is spanned by the vector fields
\begin{eqnarray*}
X_1 &=& x^1 \, \pa_1 \ + \ x^2 \, \pa_2 \ + \ w^1 \, \^\pa_1 \ + \ w^2 \, \^\pa_2 \ , \\
X_2 &=& x^2 \, \pa_1 \ - \ x^1 \, \pa_2 \ + \ w^2 \, \^\pa_1 \ - \ w^1 \, \^\pa_2\ ; \\
Y_1 &=& \exp [\la t] \, \pa_1 \ , \\
Y_2 &=& \exp [\la t] \, \pa_2 \ .  \end{eqnarray*}
Note that the $X_i$ are W-symmetries, while the $Y_i$ are deterministic symmetries.
A trivial computation shows that the commutation relations are given by
\begin{eqnarray*} & & [X_1,X_2] = 0 \ , \ \ [Y_1,Y_2] = 0 \ ; \\
& & [Y_1,X_1] = Y_1 \ , \ \ [Y_1,X_2] = - Y_2 \ , \\
& & [Y_2,X_1] = Y_2 \ , \ \ [Y_2,X_2] = Y_1 \ .  \end{eqnarray*}
This shows that $\mathcal{G}$ is indeed a Lie algebra; moreover $\G_X = \{ X_1,X_2\}$ and $\G_Y = \{ Y_1 , Y_2 \}$ are abelian subalgebras, and $\G_Y$ an abelian ideal in $\G$. \EOE

\section{The method of invariants}
\label{sec:invar}

Consider a general function \beq F \ = \ F (x,t,w) \ ; \eeq its
Ito differential is
$$ d F \ = \ \frac{\pa F}{\pa t} \, d t \ + \ \frac{\pa F}{\pa
x^i} \, d x^i \ + \ \frac{\pa F}{\pa w^a} \, d w^a \ + \ \frac12
\, \Delta F \, d t \ . $$ When we evaluate this on the dynamics
described by the Ito Eq \eqref{eq:Ito}, we get \beql{eq:dFdyn}
d F \ = \ \[ \frac{\pa F}{\pa t} \ + \ f^i (x,t) \, \frac{\pa
F}{\pa x^i} \ + \ \frac12 \, \Delta F \] \, d t \ + \
\[ \frac{\pa F}{\pa w^a} \ + \ \s^{i}_{\ a} (x,t) \, \frac{\pa
F}{\pa x^i} \] \, d w^a \ . \eeq It follows immediately that if
(and only if) $F$ satisfies the equations
\begin{eqnarray}
\frac{\pa F}{\pa t} \ + \ f^i (x,t) \, \frac{\pa F}{\pa x^i} \ + \
\frac12 \, \Delta F &=& 0 \ , \label{eq:inveq1} \\
\frac{\pa F}{\pa w^a} \ + \ \s^{i}_{\ a} (x,t) \, \frac{\pa F}{\pa
x^i} & = & 0 \ \ \ \ (a = 1,...,m) \ , \label{eq:inveq2}
\end{eqnarray} then $F$ is an \emph{invariant} for the SDEs
\eqref{eq:Ito}.

Note that in general the invariants will be arbitrary functions of some ``basic invariants'' functions.

\medskip\noindent
{\bf Example \ref{sec:invar}.1.} Consider the equation
\beql{eq:invex1} d x^i \ = \ f^i (t) \, d t \ + \ \s^i_{\ k} (t)  \, d w^k \ \ \ \ (i=1,...,n)
\ ; \eeq then the Eq \eqref{eq:inveq2} yield immediately that
$$ F(x,t,w) \ = \ F(z,t) \ ; \ \ \ z^i \ := \ x^i \ - \ \int \s^i_{\ k} (t)  \, d w^k (t)  \ . $$ A straightforward computation shows that in this case $\Delta F = 0$ (note in particular $\Delta z^i = 0$). The Eq~\eqref{eq:inveq1} reads then
$$ \frac{\pa F}{\pa t} \ - \ f^i \, \frac{\pa F}{\pa z^i} \ = \ 0 \ , $$ and it follows immediately that
$$ F \ = \ F ( \zeta ) \ , $$ having defined
\beql{eq:zetainvex1} \zeta^i \ := \ z^i \ - \ \Phi^i (t) \ = \ x^i \ - \ \Phi^i (t) \ - \ \int \s^i_{\ k} (t) \, d w^k  \ , \eeq
where $\Phi$ is the primitive of $f$,
\beql{eq:Phiexa61} \Phi^i (t) \ = \ \int \ f^i (\tau) \, d \tau \ . \eeq
Obviously, the $\zeta^i$ are in this case the basic invariants mentioned above.
\EOE

The existence of invariants allows to express the solutions to a
SDE system in terms of a reduced system of Ito equations; if there
are as many independent invariants as degrees of freedom, then the
solution can be expressed purely in terms of invariants.

In other words, the presence of a sufficient number of invariants
guarantees the integrability of the equation, pretty much as in the deterministic case the presence of a sufficient number of conserved quantities guarantees integrability.

\medskip\noindent
{\bf Example \ref{sec:invar}.2.} In the case of Eq \eqref{eq:invex1}, we have the basic invariants defined in \eqref{eq:zetainvex1}; at the initial time $t=0$ these will satisfy $\zeta^i = \zeta^i (0)$. If $w^i (0) = 0$, and choosing the constant in the primitives of \eqref{eq:Phiexa61} so that $\Phi^i (0)= 0$, at $t=0$ we have
$$ \zeta^i (0) \ = \ \zeta^i_0 \ = \ x^i_0 \ = \ x^i (0) \ . $$
Thus we have at all times and for each realization of the Wiener processes $w^k (t)$, and with $\Phi^i$ as in \eqref{eq:Phiexa61} above,
$$  x^i (t) \ = \ x^i_0 \ + \ \Phi^i (t) \ + \ \int_0^t \s^i_{\ k} (\tau) \ d w^k (\tau)   \ . $$
This yields directly the solution to \eqref{eq:invex1}. \EOE

It is natural to wonder if and how invariants are related to symmetries of the stochastic equation.

Comparing Eqs \eqref{eq:deteqIto1}, \eqref{eq:deteqIto2} one the one hand, and \eqref{eq:inveq1}, \eqref{eq:inveq2} on the other, we immediately observe a rather trivial (but useful) relation: if both $f^i$ and $\s^i_{\ k}$ do not actually depend on the $x$ variables (but possibly depend on $t$) and set $R=0$, the symmetry coefficients $\vphi^i$ are also invariants for the Ito equation. (Note that such relation does not exist for $R \not=0$, i.e., for proper W-symmetries \gcite{GW18}.)

We would be interested in knowing if other relations exist. This matter has been studied in recent papers by one of us \gcite{Koz18a,Koz18b,KozB} (to which we refer for details and examples), and we briefly report the relevant results here.

\medskip\noindent
{\bf Proposition \ref{sec:invar}.1.} {\it The set of invariants for a given Ito equation is a ring over ${\bf R}$, i.e., if $F,G$ are invariants for the given equations and $\a,\b$ are real constants, so are
$$ H \ = \ \a \, F \ + \ \b \, G \ , \ \ \ K \ = \ F \cdot G \ . $$}

\medskip\noindent
{\bf Proposition \ref{sec:invar}.2.} {\it If a given Ito Eq \eqref{eq:Ito} admits the symmetry $X$ and the invariant $F$, then $G = X(F)$ is also an invariant for the same equation.}

\medskip\noindent
{\bf Proposition \ref{sec:invar}.3.} {\it The set of symmetries for a given Ito equation has, beside the structure of Lie algebra, the structure of a  \emph{Lie module} over the ring of invariants; i.e., if $X,Y$ are symmetries for the given equation, and $F,G$ invariants for the same equation, then
$$ Z \ = \ F \, X \ + \ G \, Y $$
is still a symmetry for the equation.}

\medskip\noindent
{\em Proof.} Proposition \ref{sec:invar}.1 is trivial; Propositions \ref{sec:invar}.2 and \ref{sec:invar}.3 are respectively Theorems 4.4 and 4.6 in \gcite{Koz18a}; see there for proofs. \EOP

We stress that, obviously, a generic stochastic equation has no invariants, even in the case it has symmetries.

\medskip\noindent
{\bf Example \ref{sec:invar}.3.}
Consider again the two-dimensional isotropic linear stochastic Eq \eqref{eq:2DISO} considered in Example \ref{sec:symmsto}.6 above (where it was found it has a nontrivial symmetry algebra). Now a direct computation shows that there are no invariants. In fact, invariants should satisfy the equations
\begin{eqnarray}
& &  J_{w_1} \, + \, \mu \, J_{x1}  \ = \ 0 \ , \nonumber \\
& &  J_{w_2} \, + \, \mu \, J_{x_2}  \ = \ 0 \ ; \label{eq:ex53} \\
& &  J_t \, + \, \la \, \( x_1 \, J_{x_1} \, + \, x_2 \, J_{x_2} \)  \ = \ 0 \ . \nonumber \end{eqnarray}
Note in the last equation there should also be a term $(1/2) \Delta J$ with $\Delta$ the Ito Laplacian \eqref{eq:ItoLap}, but in this case it results $\Delta J = 0$.) Solving the first two equations yields
$$ J (x_1,x_2,t,w_1,w_2) \ = \ H (z_1,z_2,t) \ ; \ \ \ z_i \ := \ w_i \ - \ x_i / \mu \ . $$
Plugging this into the last Eq \eqref{eq:ex53} yields $H = const$, thus showing that only trivial invariants are present. \EOE


\section{Conditional and asymptotic symmetries for stochastic equations}
\label{sec:cassto}

We now want to argue that -- pretty much in the same way as a
great extent of the standard theory of symmetry of deterministic
differential equations can be extended to the framework of
stochastic differential equations -- the concepts of conditional,
partial and asymptotic symmetry also apply for stochastic
differential equations. What is more, these are also \emph{useful} to
determine solutions to stochastic equations.

It will happen that these are intimately related -- much more than in the deterministic setting -- to similar concepts for invariants, i.e., to conditional, partial or asymptotic invariants. These, and the relation with symmetries, will be discussed in the next Sect. \ref{sec:caisto}. We will first discuss the matter in abstract terms -- actually with a more formal writing than in previous Sections, as now we develop new material and results; and then provide, in Sect. \ref{sec:examples} below, a number of concrete Examples.


\subsection{Conditional and partial symmetries}
\label{sec:CPSsto}

Our first observation is that conditional and partial symmetries can be, in principle, defined for stochastic equations exactly as in the case of deterministic ones.

The difference -- and the reason to write ``in principle'' in the lines above -- is that while a solution of a deterministic equation is determined by the initial condition, in the case of stochastic equation it is determined by the initial condition \emph{and} by the realization of the driving Wiener processes. But, of course, we do not know a priori what the realization of the Wiener process will be, so we do not want to discuss properties which depend on such realization. (A partial exception to this is that often we are satisfied with properties which hold for \emph{generic} realizations of the Wiener processes, i.e., for a full measure set of the driving Wiener paths.)

Thus, in practice, it is quite difficult -- and makes little sense in view of applications -- to study symmetries of a given solution, or of a given set of solutions.

The exception is provided by the case where the solutions in the set which is studied are identified by their living on an \emph{invariant manifold} for the equation. By this we mean that the Eq \eqref{eq:Ito} describes a variable $(x,t) \in M$ ($M$ is the phase manifold), and there is a proper submanifold $M_0 \ss M$ such that if $(x(0),t_0) \in M_0$, then \emph{for any realization} of the driving Wiener processes, and for any $t \ge t_0$, $(x(t),t) \in M_0 $.

We will actually confine ourselves to the case of \emph{autonomous} equations, i.e., the case where the coefficients $f^i (x,t)$ and $\s^i_{\ k} (x,t)$ in \eqref{eq:Ito} are actually independent of $t$. In this case we can work with the reduced phase manifold\footnote{This was denoted as $M_0$ in Sect. \ref{sec:symm}; but here we reserve this notation to the dynamically invariant submanifold. We trust no confusion will arise.}, and in the following $M$ should be interpreted as the reduced phase manifold. Extension of our discussion and results to the non-autonomous is rather simple, and in any case can be obtained -- remaining within the autonomous formalism -- by adding a new variable $x^0$ with evolution governed by $d x^0 = d t$.

We will see in the following that the existence of these invariant manifolds is naturally related to ``conditional'' (or ``partial'') invariants for the equation.\footnote{The simplest -- but rather trivial -- case in which there is an invariant manifold is that of a \emph{fixed point}; for example the stochastic Eq~\eqref{eq:Ito} with coefficients satisfying $f^i (0,t)=0$, $\s^i_{\ j} (0,t) = 0$, automatically admit $\xb = 0$ as a fixed point, and hence a (highly symmetric) solution $x(t) \equiv 0 $. This is of little interest.}

The simplest nontrivial case is that in which the dynamics admits a nontrivial
invariant submanifold $M_0 \ss M$ (this is of course more significant if
$M_0$ is stable or attracting, as will be discussed later on). In this case we can consider the
restriction of our system to $M_0$, and this may have symmetries
which are not present for the full system. These in turn may allow
for the integration of the restricted equation; solutions obtained
in this way are also special solutions for the full system.

We start by stating some very simple -- but useful --  facts.

\medskip\noindent
{\bf Lemma \ref{sec:cassto}.1.} {\it Let the SDE $\Delta$
\eqref{eq:Ito} in the $n$-dimensional manifold $M$ admit an invariant submanifold $M_0 \ss M$ of dimension $m < n$. Consider the restriction $\De_0$ of
$\De$ to $M_0$. Any solution to $\De_0$ is also a solution to $\De$, and the most general solution to $\De$ with $x (0) \in M_0$ is obtained as a solution to $\De_0$.}

\medskip\noindent
{\em Proof.} Obvious.

\medskip\noindent
{\bf Lemma \ref{sec:cassto}.2.} {\it In the setting of Lemma \ref{sec:cassto}.1, let $\De_0$ admit $m$ simple symmetries spanning a solvable algebra. Then it can be integrated, and hence all the solutions to the initial equation $\De$ with initial
conditions on $M_0$ can be obtained.}
\bigskip

\medskip\noindent
{\em Proof.} Integrability of $\De_0$ follows from  Proposition \ref{sec:symmsto}.3 above. Once $\De_0$ is integrable, the rest of the statement follows immediately from Lemma \ref{sec:cassto}.1. \EOP

This construction is also generalized to the case where the symmetry algebra has smaller dimension, as stated in the following Lemma.

\medskip\noindent
{\bf Lemma \ref{sec:cassto}.3.} {\it Let the Ito SDE $\Delta$ in the $n$-dimensional manifold $M$ admit an invariant submanifold $M_0 \ss M$ of dimension $m < n$. Let $\De_0$ be the restriction of $\De$ to $M_0$. Then \par\noindent $(a)$ If $\De_0$ admits $s<m$ simple symmetries spanning a solvable algebra $\G$ which generates a local Lie group $G$, then
$\De_0$ can be reduced to an equation $\^\De_0$ on $M_0/G$.
\par\noindent $(b)$ Let $\^x (t)$ be any solution to $\^\De_0$; then $\^x (t)$ extends to a solution $x(t)$ of $\De_0$, and this is turn describes the most general solution to $\De$ with $x(0) \in M_0$.}

\medskip\noindent
{\em Proof.} Point $(a)$ follows by Proposition \ref{sec:symmsto}.3 above. Solutions to $\^\De_0$ are extended to solutions to $\De_0$ again by means of Proposition \ref{sec:symmsto}.3, i.e., via the reconstruction equations. In view of Lemma \ref{sec:cassto}.1, the exact solutions obtained in this way are also all the solutions to the initial equation $\De$ with initial conditions on $M_0$. \EOP

\medskip\noindent
{\bf Remark \ref{sec:cassto}.1.} The statement in these Lemmas are rather trivial, as shown by their very simple proofs. Actually these Lemma are more the description of a strategy to obtain a set of solutions (identified by living on $M_0$) to $\De$. \EOR

\medskip\noindent
{\bf Remark \ref{sec:cassto}.2.} If the equation admits a standard (global) invariant, then all the level manifolds for this invariant are invariant manifolds of the differential equation; see Example \ref{sec:examples}.1 below. Thus the Lemmas \ref{sec:cassto}.1, \ref{sec:cassto}.2 and \ref{sec:cassto}.3 given above apply in general, in the sense that \emph{any}  initial datum lies on such an invariant manifold. But they apply also to the case where no global invariant exists (we will see in a moment a weaker type of invariant is involved here), i.e., when we have \emph{isolated} invariant manifolds. \EOR

\medskip\noindent
{\bf Remark \ref{sec:cassto}.3.} The presence of an isolated invariant manifold can be described in terms of a \emph{conditional invariant} for the stochastic equation. That is, if $M_0$ is the submanifold described by
$F (x) = m_0$ (here $m_0 \in {\bf R}$) and $M_0$ is an invariant manifold while there are no nearby invariant manifolds, then $F (x)$ is a conditional invariant for our equation. Note that this means that on the dynamics we have
$$ d F \ = \ [ F(x) - m_0 ] \ \[ \a (x,t) \, d t \ + \ \b_k (x,t) \, d w^k \] \ , $$ with nonzero $\a, \b_k$. In this case \emph{only} the $m_0$ level set is invariant, while the other ones are in general not invariant, as in our Examples \ref{sec:examples}.2 and \ref{sec:examples}.3 below (in both cases the conditional invariant is just $\rho$ and the invariant level set is $\rho = 1$). We will discuss this point in more detail in Sect. \ref{sec:caisto} below. \EOR

\medskip\noindent
{\bf Remark \ref{sec:cassto}.4.} This situation is similar to the setting considered by Misawa and by Albeverio \& Fei \gcite{AlbFei,Mis1,Mis2,Mis3} in their early study of symmetry of stochastic equations (see the discussion in \gcite{GGPR}). In their case, however, we had a global (proper) invariant, hence a global constant of motion -- which allowed for a dimensional reduction of the stochastic system -- as in Example \ref{sec:examples}.1 below, while in the setting discussed here we only have a conditional one, hence reduction can take place \emph{only} on the specific level set of the conditional constant of motion, viz. the manifold $M_0$. \EOR

\medskip\noindent
{\bf Remark \ref{sec:cassto}.5.} The situation discussed in this Section is related to \emph{conditional constants of motion} \gcite{SLC,PucRos1,PucRos2,Hall1,Hall2} and \emph{conditional invariants}, so that it would be natural to speak of conditional symmetries. On the other hand, in this case the solutions (to the restricted problem) are \emph{not} invariant under the conditional symmetry: they are actually mapped one into another by the action of the latter. In this sense, the situation is more similar to that encountered when dealing, in the deterministic framework, with \emph{partial} symmetries; from this point of view it might be more appropriate to speak of \emph{partial invariants}; but we will prefer to use the term \emph{conditional invariant} in view of the first mentioned parallel with the deterministic case, and the established use in the latter. \EOR

\subsection{Invariant manifolds and (conditional) configurational invariants}
\label{sec:IMCCI}

In the previous discussion, in particular Lemmas \ref{sec:cassto}.1 through \ref{sec:cassto}.3, we have assumed the existence of an invariant manifold $M_0 \ss M$ for the dynamics described by our Ito Eq \eqref{eq:Ito}. Needless to say, this is a highly non-generic situation. In fact, as anticipated in Remark \ref{sec:cassto}.3, the existence of invariant manifolds is related to the existence of (conditional) invariants for the dynamics.

In order to identify invariant manifolds in $M$, or in $M \times {\bf R}$, these invariants should depend only on the $x$, or the $(x,t)$, variables. We have thus the following:

\medskip\noindent
{\bf Definition \ref{sec:cassto}.1.} {\it When the invariant $J = J(x,t;w)$ for an Ito equation does not depend on the $w$ variables, i.e., $J(x,t;w) = \Psi (x,t)$, we say that $J$ is a \emph{phase invariant}; if moreover $\Psi$ is independent of $t$, we say $J$ is a \emph{configurational invariant}.}

\medskip\noindent
{\bf Remark \ref{sec:cassto}.6.} We stress that while configurational invariants can identify invariant manifolds, and phase invariant can identify time-varying invariant manifolds, in order to identify solutions in terms of invariants we need invariants which also depend on the $w^i$ variables, and hence on the realization of the Wiener processes. See also Example \ref{sec:examples}.1 in this respect. \EOR

We recall that if $J : M \to R$ is a smooth function defined on the phase manifold $M$, or more generally $J : M \times \mathcal{W} \to R$ where $\mathcal{W} \approx {\bf R}^n$ is the space in which the driving Wiener processes $w^i = w^i (t)$ take values, then its evolution on the dynamics described by the Ito equation $\De$ -- written as \eqref{eq:Ito} -- is given by
\begin{eqnarray}
d J &=& \( \frac{\pa J}{\pa t} \) \, dt \ + \ \( \frac{\pa J}{\pa x^i} \) \, d x^i \ + \ \frac12 \ \Delta (J) \, d t \nonumber \\
&=& \[ \( \frac{\pa J}{\pa t} \) \ + \ \( \frac{\pa J}{\pa x^i} \) \ f^i (x,t) \ + \ \frac12 \ \Delta (J) \] \ dt \label{eq:dJfull} \\
& & \ \ + \ \[ \( \frac{\pa J}{\pa w^k} \) \ + \ \( \frac{\pa J}{\pa x^i} \) \, \s^i_{\ k} (x,t) \] \ d w^k \ . \nonumber  \end{eqnarray}

We are specially interested in the case where $J$ is a function of $x$ and $t$ alone, $J(x,t;w) = \Psi (x,t)$; we say then that $J$ is a \emph{phase space function}, and if $\Psi$ does not depend on $t$ we say it is a \emph{configurational function}. In both these cases the Ito Laplacian reduces to
\beql{eq:deltaJpsi} \Delta J \ = \ \Delta \Psi \ = \ \s^{ik} \s^j_{\ k} \ \frac{\pa^2 \Psi}{\pa x^i \pa x^j} \ . \eeq

\medskip\noindent
{\bf Lemma \ref{sec:cassto}.4.} {\it The evolution of a phase space function under the dynamics described by an Ito Eq \eqref{eq:Ito} is itself described by an Ito equation.}

\medskip\noindent
{\em Proof.} Using \eqref{eq:dJfull} and specializing it to the case $J = \Psi (x,t)$, we have
\begin{eqnarray}
d J &=& \[ \( \frac{\pa \Psi}{\pa t} \) \ + \ \( \frac{\pa \Psi}{\pa x^i} \) \ f^i (x,t) \ + \ \frac12 \ \Delta (\Psi) \] \ dt \nonumber \\ & & \ + \ \[ \( \frac{\pa \Psi}{\pa x^i} \) \, \s^i_{\ k} (x,t) \] \ d w^k \ . \label{eq:dJx}  \end{eqnarray}
Recalling also \eqref{eq:deltaJpsi}, we note that now all terms within the square brackets depend on $(x,t)$ alone, i.e., we have again an equation of Ito type. \EOP

It is now natural to introduce the following Definition, which leads immediately to the foregoing Lemma \ref{sec:cassto}.5.

\medskip\noindent
{\bf Definition \ref{sec:cassto}.2.} {\it \emph{(a)} If the function $J$ is such that $d J = 0$ on solutions to the Ito equation \eqref{eq:Ito}, i.e., \eqref{eq:dJfull} is identically zero, we say that it is an \emph{invariant} for the Ito equation (a \emph{phase space invariant} if $J = \Psi (x,t)$, and \eqref{eq:dJx0} is identically satisfied, a \emph{configurational invariant} if $J = \Psi (x)$).
\par\noindent
\emph{(b)}
If this relation is valid only on the level set $J = c$, we say that $J$ is a \emph{conditional invariant} (a \emph{conditional phase space invariant} if $J = \Psi (x,t)$, a \emph{conditional configurational invariant} if $J = \Psi (x)$) for the Ito equation.}

\medskip\noindent
{\bf Lemma \ref{sec:cassto}.5.} {\it Phase space invariants are identified by the $n+1$ equations
\begin{eqnarray}
& & \( \frac{\pa \Psi}{\pa t} \) \ + \ \( \frac{\pa \Psi}{\pa x^i} \) \ f^i (x,t) \ + \ \frac12 \ \Delta (\Psi)  \ = \ 0 \ , \nonumber \\
& & \( \frac{\pa \Psi}{\pa x^i} \) \, \s^i_{\ k} (x,t)  \ = \ 0 \ . \label{eq:dJx0}  \end{eqnarray}}

\medskip\noindent
{\em Proof.} This follows at once from \eqref{eq:dJx}. \EOP

\medskip\noindent
{\bf Remark \ref{sec:cassto}.7.}
If the solutions to the \eqref{eq:dJx0} identify one or more level sets of $J=\Psi$ (in practice, if the l.h.s. of the above equations can be factorized with a factor being just a function of $\Psi$), then this or these level sets correspond to invariant manifolds for the dynamics. We will see in concrete Examples that this may happen. \EOR

\medskip\noindent
{\bf Remark \ref{sec:cassto}.8.} As configurational invariants (and functions) are special cases of phase space invariants (and functions), the Lemmas \ref{sec:cassto}.4 and \ref{sec:cassto}.5 above also apply to them. Note that the Eq \eqref{eq:dJx0} read in this case
\begin{eqnarray}
& & \( \frac{\pa \Psi}{\pa x^i} \) \ f^i (x) \ + \ \frac12 \ \Delta (\Psi)  \ = \ 0 \ , \nonumber \\
& & \( \frac{\pa \Psi}{\pa x^i} \) \, \s^i_{\ k} (x)  \ = \ 0 \ , \label{eq:dJx00}  \end{eqnarray}
where we have taken into account the fact that configurational invariants naturally arise in autonomous systems.
\EOR

\subsection{Asymptotic symmetries}
\label{sec:asysymsto}

We have so far discussed the case where an invariant submanifold $M_0 \ss M$ exists, and considered dynamics on it. We have not discussed neither what happens nearby the invariant manifold, nor the stability of such manifold.

As we are considering stochastic systems, we should consider at least two types of (local or global) stability \gcite{Kushner,Khasminski}.

\begin{itemize}
\item[(A)] On the one hand, it is possible that for any initial datum $x_0$ in a neighborhood $\mathcal{B}_0 \supset M_0$ and any realization of the Wiener processes $w^i (t)$, the dynamics is attracted towards $M_0$.
\item[(B)] On the other hand, it is possible that for any $x_0 \in \mathcal{B}_0$, considering the Fokker-Planck equation with initial condition $u (x,0) = \delta (x - x_0)$ the probability density function $u(x,t)$ remains concentrated around $M_0$.
\end{itemize}

\medskip\noindent
{\bf Definition \ref{sec:cassto}.3.} {\it When the situation described in case (A) above occur, we say that $M_0$ is locally \emph{strongly attractive}; if this holds for any $x_0 \in M$, we say it is \emph{globally strongly attractive}.}

\medskip\noindent
{\bf Definition \ref{sec:cassto}.4.} {\it When the situation described in case (B) above occur, we say that $M_0$ is locally \emph{weakly attractive}; if this holds for any $x_0 \in M$, we say it is \emph{globally weakly attractive}.}

\medskip\noindent
{\bf Remark \ref{sec:cassto}.9.} Note that in case (B) above typically there will be a parameter $\s_0 \ge 0$ -- related in physical terms to the intensity of the noise described by the Wiener processes -- which controls the spreading of the asymptotic measure around $M_0$, and such that in the limit $\s_0 \to 0$ the asymptotic measure $u_* (x) = \lim_{t \to \infty} u(x,t)$ collapses to a measure on $M_0$. \EOR

\medskip\noindent
{\bf Lemma \ref{sec:cassto}.6.} {\it In the case (A), the most general \emph{asymptotic} solution to $\De$ will be described by the most general solution to $\De_0$.}

\medskip\noindent
{\em Proof.} In fact, under these hypotheses any dynamics starting in $\mathcal{B}$ will be attracted to the invariant manifold $M_0$, and on this the dynamics is described by $\De_0$. The conclusion follows then by Lemma \ref{sec:cassto}.1 above. \EOP

\medskip\noindent
{\bf Remark \ref{sec:cassto}.10.} Needless to say, even in the case $\De_0$ is fully integrable we will in general not be able to describe which ones of the solutions to $\De_0$ will describe the asymptotic behavior for a given solution to $\De$, i.e., for given initial datum $x_0 \in \mathcal{B}$ and given realization of the Wiener processes, as we will not be able to integrate the transient dynamics. \EOR

\medskip\noindent
{\bf Remark \ref{sec:cassto}.11.} The Remark above refers to case (A). In case (B), on the other hand, the dynamics can not be reduced, even asymptotically, \emph{exactly} to the dynamics on $M_0$ and hence to $\De_0$. Thus even in the case where $\De_0$ is exactly integrable -- and even for $x_0 \in \mathcal{B}_0$ -- this does not give a description of the asymptotic dynamics for $\De$. In this case we have a weaker result. \EOR

\medskip\noindent
{\bf Lemma \ref{sec:cassto}.7.} {\it In the case (B), for small enough $\s_0$, the asymptotic dynamics will be described for most of the time, by the \emph{linearization} $\wt{\De}_0$ of $\De$ around $M_0$.}

\medskip\noindent
{\em Proof.} Under the hypotheses holding in case (B) and for small enough $\s_0$, the measure will be asymptotically concentrated around $M_0$, so that the linearization  $\wt{\De}_0$ (which, in turn, will be a linearization around $\De_0$) will describe ``most'' of the asymptotic solutions; by this we mean that there could be from time to time \emph{large fluctuations} driving the dynamics for a short time outside of the immediate neighborhood of $M_0$, i.e., outside the region where $\De$ reduces to $\wt{\De}_0$. These large fluctuations will occur with a frequency exponentially small in $\s_0$ and will decay rapidly so that the dynamics is quickly driven back to the region around $M_0$. \EOP

\medskip\noindent
{\bf Remark \ref{sec:cassto}.12.} This Lemma does not give a result of the same strength as the previous one; on the other hand, it says that one will have a reliable description in terms of $\wt{\De}_0$, except for rare and short fluctuations. \EOR

As far as symmetries are concerned, our discussion suggests naturally the following definitions.

\medskip\noindent
{\bf Definition \ref{sec:cassto}.5.} {\it Let $M_0 \ss M$ be a strongly attractive submanifold for the equation $\De$, and let $\De_0$ be the restriction of $\De$ to $M_0$; let $X$ be a symmetry for $\De_0$, hence a conditional symmetry for $\De$. We say that $X$ is also a \emph{strong asymptotic symmetry} for $\De$.}

\medskip\noindent
{\bf Definition \ref{sec:cassto}.6.} {\it Let $M_0 \ss M$ be a weakly attractive submanifold for the equation $\De$, let $\De_0$ be the restriction of $\De$ to $M_0$, and let $\wt{\De}_0$ be the linearization of $\De$ around $M_0$; let $X$ be a symmetry for $\De_0$, hence a conditional symmetry for $\De$. \emph{If} $X$ is also a symmetry for $\wt{\De}_0$, then we say that $X$ is also a \emph{weak asymptotic symmetry} for $\De$.}

Both these cases are related to the presence of \emph{asymptotic} conditional invariants, as we are now going to discuss in the next Section \ref{sec:caisto}.

We stress that we are again interested in the case of phase space or configurational invariants, i.e., where the would-be invariants depend on $(x,t)$ or on the $x$ variables alone; thus we are in the setting discussed above, and the evolution is described by \eqref{eq:dJx}.

\section{Asymptotic invariants and asymptotic symmetries for stochastic equations}
\label{sec:caisto}

The discussion of Section \ref{sec:IMCCI}, referring to conditional symmetries, can be extended to the case of asymptotic symmetries.

The key observation is that the evolution of a phase space function under the dynamics of an Ito Eq \eqref{eq:Ito} is described by \eqref{eq:dJx}, i.e., by an equation of Ito type; this will be denoted, for ease of reference, as $\De_J$. So our discussion of (stochastic) stability for submanifolds $M_0$ of $M$ under $\De$ can be repeated for the stability under $\De_J$.

\medskip\noindent
{\bf Definition \ref{sec:caisto}.1.} {\it If $J = \Psi (x)$ is a conditional configurational invariant for the Ito equation $\De$ on the invariant manifold $M_0$, and $M_0$ is a strongly attractive manifold for $\De$, we say that $J$ is a \emph{strong asymptotic invariant} for $\De$.}

\medskip\noindent
{\bf Definition \ref{sec:caisto}.2.} {\it If $J = \Psi (x)$ is a conditional configurational invariant for the Ito equation $\De$ on the invariant manifold $M_0$, and $M_0$ is a weakly attractive manifold for $\De$, se say that $J$ is a \emph{weak asymptotic invariant} for $\De$.}

\medskip\noindent
{\bf Lemma \ref{sec:caisto}.1.} Let $J$ be a strongly asymptotic configurational invariant for the Ito equation $\De$, associated to the invariant manifold $M_0$, and let $\De_0$ be the reduction of $\De$ to $M_0$. Then the asymptotic solutions to $\De$ are described by the solutions to $\De_0$.

\medskip\noindent
{\em Proof.} By definition, under the hypotheses of the Lemma $M_0$ is a strongly attractive invariant manifold for $\De$, hence the statement follows immediately from Lemma \ref{sec:cassto}.6, which itself uses Lemma \ref{sec:cassto}.1. \EOP

\medskip\noindent
{\bf Lemma \ref{sec:caisto}.2.} Let $J$ be a weakly asymptotic configurational invariant for the Ito equation $\De$, associated to the invariant manifold $M_0$, and let $\wt{\De}_0$ be the linearization of $\De$ around $M_0$. Then the asymptotic solutions to $\De$ are described by the solutions to $\wt{\De}_0$, up to large fluctuations.

\medskip\noindent
{\em Proof.} By definition, under the hypotheses of the Lemma $M_0$ is a weakly attractive manifold for $\De$. Thus, according to Lemma \ref{sec:cassto}.7, the dynamics of $\De$ lives, for most of the time and up to rare and short-lived large fluctuations, in a neighborhood of $M_0$; thus it is described by $\wt{\De}_0$, and the Lemma is just a restatement of these facts in terms of (asymptotic) invariants. \EOP

\medskip\noindent
{\bf Remark \ref{sec:caisto}.1.} The approach discussed in this subsection is strongly related to stochastic Lyapounov functions, see e.g., \gcite{Kushner,Khasminski}. Some of our ideas are also related to ideas developed in the context of bifurcation for stochastic systems \gcite{Doan}. \EOR

\medskip\noindent
{\bf Remark \ref{sec:caisto}.2.} Our discussion used \emph{configurational} invariants to relate invariants to attractive invariant submanifolds. As stressed above, see Remark \ref{sec:cassto}.6, configurational or phase space (asymptotic) invariants can only do this, i.e., they cannot be used to obtain solutions to Ito equations, at difference with general (asymptotic) invariants, i.e., invariants depending also on the $w$ variables. See also the discussion in Sect. \ref{sec:invar} for the case of full invariants. \EOR


\section{Examples}
\label{sec:examples}

We will now consider concrete Examples related to the matters discussed in Sections \ref{sec:cassto} and \ref{sec:caisto} above. These matters are of course strongly interconnected, but we have separated the Examples in different types -- and correspondingly separated the present Section in different subsections -- for ease of reading.

In most cases, our Examples are built so that the compact invariant manifold is just a circle (usually the unit one), as we want to focus on conceptual issues rather than on computational difficulties. Thus it is not surprising that things will be simpler in polar or spherical coordinates, and indeed we will mostly work in polar coordinates, albeit in some cases we will also start from Cartesian ones.

\subsection{Full symmetry, full invariants}

We start by considering an Example due to Misawa \gcite{Mis3}; this is Example 1 in there, where it is considered in the equivalent Stratonovich form, and only up to checking $J$ is an invariant.

\medskip\noindent
{\bf Example \ref{sec:examples}.1.}
Consider the Ito equation
\begin{eqnarray}
d x_1 &=& \( x_3 \, - \, x_2 \, - \, \frac12 \, ( 2 x_1 - x_2 - x_3 ) \) \, d t \ + \ (x_3 \, - \, x_2 ) \, d w \nonumber \\
d x_2 &=& \( x_1 \, - \, x_3 \, - \, \frac12 \, ( 2 x_2 - x_3 - x_1 )\) \, d t \ + \ (x_1 \, - \, x_3 ) \, d w \label{eq:Misawa} \\
d x_3 &=& \( x_2 \, - \, x_1 \, - \, \frac12 \, ( 2 x_3 - x_1 - x_2 ) \) \, d t \ + \ (x_2 \, - \, x_1 ) \, d w \nonumber \end{eqnarray}
(note that only one Wiener process is involved). In this case the function
\beq  J \ = \ x_1^2 \, + \, x_2^2 \, + \, x_3^2  \ = \ \rho^2  \eeq is an invariant. In fact one has
$$ \Delta J \ = \ 4 \ \[ J \ - \ \( x_1 x_2 + x_2 x_3 + x_3 x_1 \) \] \ , $$
and then $d J = 0 $ on the Eq \eqref{eq:Misawa} follows immediately from a direct computation. Thus any sphere of radius $\rho$ is invariant under the evolution described by \eqref{eq:Misawa}.

It is also easily checked, by the same procedure or by just looking at \eqref{eq:Misawa}, that
\beq H \ = \ x_1 \ + \ x_2 \ + \ x_3 \eeq is another invariant.

By considering the Jacobian matrix for the change of coordinates $(x_1,x_2,x_3) \to (J,H,z)$ with $z = x_3$, we observe that its determinant is $2 (x_1 - x_2)$, i.e., this is singular on the plane $x_1 = x_2$; we will thus have to consider two regions.

Implementing the change of variables $(x_1,x_2,x_3) \to (J,H,z) $ we have that the inverse change is given (in the two regions $x_1 < x_2$ and $x_1 > x_2$) by
\begin{eqnarray*}
x_1 &=& \frac12 \ \[ ( H \ - \ z ) \ \mp \ \sqrt{2 (J - z^2 ) \ - \ (H - z )^2} \] \ , \nonumber \\
x_2 &=& \frac12 \ \[ ( H \ - \ z ) \ \pm \ \sqrt{2 (J - z^2 ) \ - \ (H - z )^2} \] \ .  \end{eqnarray*}
Note that by elementary algebra, and in view of the definitions of $J,H$, and $z$, we have $$ 2 (J - z^2 ) \ - \ (H - z )^2 \ = \ (x_1 \ - \ x_2 )^2 \ ; $$
thus the argument of the square root in the formula above is always positive. We introduce, for ease of writing, the new \emph{real} function
$$ \chi (J,H;z) \ = \ \sqrt{2 (J - z^2 ) \ - \ (H - z )^2 } \ , $$ so that the above inverse change of variables reads
\begin{eqnarray}
x_1 &=& \frac12 \ \[ ( H \ - \ z ) \ \mp \ \chi (J,H;z) \] \ , \nonumber \\
x_2 &=& \frac12 \ \[ ( H \ - \ z ) \ \pm \ \chi (J,H;z) \] \ . \label{eq:MisCV} \end{eqnarray}

On the one-dimensional manifolds (corresponding to the $\pm$ sign in the formulas above) identified by given values for $J = \rho^2$ and $H$ we have the reduced equation
\beq dz \ = \ \[  \frac12 \ (H - 3 z) \ \pm \ \chi (J,H;z) \] \ d t \ \pm \ \chi (J,H;z) \ d w  \ . \eeq
Note that here $z \in [-\rho,\rho]$ (recall $J = \rho^2$), and that for $z = \pm \rho$ we have $J = z^2$, $H = z$ and $x_1=x_2 = 0$.

We can further look for symmetries of the reduced equation. We disregard W-symmetries, i.e., set $R=0$ in \eqref{eq:deteqIto2}\footnote{The case with $R\not= 0$ can also be analyzed completely, yielding of course more complex formulas.}; with straightforward computations we get
$$ \vphi_\pm (z,t;w) \ = \ \chi (J,H;z)  \ \psi_\pm ( u_\pm ) \ , $$
where the $\pm$ refers to the two determinations given above for $x_1$ and $x_2$ in terms of $z$, $\psi_\pm$  are arbitrary function of their argument, and
$$ u_\pm \ = \ w \ + \ t \ \pm \ \frac{1}{\sqrt{3}} \ \arctan \[ \frac{H - 3 z}{\sqrt{3} \, \chi (J,H;z)} \] \ . $$
The simplest case does of course correspond to $\psi_\pm ( u_\pm  ) = c_\pm $, say $$ \psi_\pm ( u_\pm ) = \pm 1 $$ for ease of discussion (note the two $\pm$ signs are independent),  i.e., to\footnote{This notation might be confusing: a more precise notation would be $\vphi_\pm = s_\pm \chi$, with $s_\pm$ a sign.}
$$ \vphi_\pm (z,t;w) \ = \  \pm \ \chi (J,H;z) \ . $$

With this, the integrating change of variables is given by $$ \xi \ = \ \int \frac{1}{\vphi_\pm (z,t;w)} \ d z \ = \ { \mp } \ \frac{1}{\sqrt{3}} \ \arctan \[ \frac{H - 3 z}{\sqrt{3} \ \chi (J,H;z) }  \] \ . $$
With a direct application of Ito rule, the evolution equation for the random variable $\xi$ turns out to be
\beql{eq:MisXi} d \xi \ = \ d t \ + \ d w \ . \eeq
This is immediately integrated, yielding
$$ \xi (t) \ = \ \xi (t_0) \ + \ (t - t_0) \ + \ [ w(t) - w(t_0) ] \ . $$
Going back to the $z$ variable we obtain (note that $3 J - H^2 \ge 0$)
$$ z \ = \ \frac{H}{3} \ \pm \ \frac{\sqrt{2}}{3}  \, \sqrt{3 J - H^2}  \ { \sin ( \sqrt{3} \, \xi) } \ . $$ Having obtained $z(t)$, i.e., $x_3 (t)$, now $x_1 (t) $ and $x_2 (t)$ are also obtained via \eqref{eq:MisCV}.
\EOE

\subsection{Full or conditional symmetries, conditional invariants}

We now consider several examples with conditional invariants; at this stage we will not yet look at asymptotic symmetries or invariants.

\medskip\noindent
{\bf Example \ref{sec:examples}.2.} Let us work in $R^2$ with polar
coordinates $(\rho,\vartheta)$, and consider the Ito equations
\begin{eqnarray}
d \rho &=&  a \, (1 - \rho^2) \, \rho \, d t \ + \ \s \, (1 - \rho^2) \, d w_1 \nonumber \\
d \vartheta &=& \[ b \, (1 - \rho^2) \ + \ \om \, \rho^2  \] \ d t
\ + \ \s \, d w_2 \label{eq:ex_6_1} \end{eqnarray} with $a,b,\om,\s$ nonzero real constants.

It is clear that the unit circle $\rho = 1$ is an invariant
manifold for these equations; the dynamics on the unit circle is
described by
$$ d \vartheta \ = \ \om \, d t \ + \ \s \, d w_2 $$
i.e., by a uniform rotation with a superimposed Wiener process.
This yields immediately
$$ \vartheta (t) \ = \ \vartheta (0) \ + \ \om \, t \ + \
\s \, [w_2 (t) - w_2 (0) ] \ ; \ \ \ \rho (t) = 1 \ .  $$
Thus albeit we have no information about the dynamics for generic
initial conditions, we have a full description of the dynamics
with initial condition on the unit circle.

The discussion is immediately generalized -- in its entirety -- to equations of the form
\begin{eqnarray*}
d \rho &=&  \a (\rho,\vartheta) \, d t \ + \ \s_1 (\rho,\vartheta) \, d w_1 \\
d \vartheta &=&  \b (\rho,\vartheta)  \ d t \ + \ \s_1 (\rho,\vartheta) d w_2 \end{eqnarray*}
with $\a,\b,\s_i$ smooth functions and \begin{eqnarray*} & & \a (\rho_0,\vartheta) \ = \ 0 \ = \ \s_1 (\rho_0,\vartheta) \ , \\ & &  \pa_\vartheta \b (\rho_0,\vartheta) \ = \ \pa_\vartheta  \s_1 (\rho_0,\vartheta) \ = \ 0 \end{eqnarray*} for some $\rho_0$.

If we drop the last condition, i.e., require only $\a (\rho_0,\vartheta) = 0 =  \s_1 (\rho_0,\vartheta)$, we still have reduction  to a one dimensional equation on the circle of radius $\rho_0$, albeit we have no information on the solutions living on this circle.

Note that in this case the full equations are rotationally invariant, as immediately seen from \eqref{eq:ex_6_1}; but the associated function $J = \rho$ is a \emph{conditional} invariant. \EOE

\medskip\noindent
{\bf Example \ref{sec:examples}.3.} Consider $R^3$ with cylindrical
coordinates $(\rho,\vartheta,z)$ and the Ito equations
\begin{eqnarray}
d \rho &=& a (1 - \rho^2) \, d t \ + \ \kappa \, (1 - \rho^2) \, d w_1 \nonumber \\
d \vartheta &=& \[ \omega \, + \, b \, (1 - \rho^2) \, + \, c \, \cos (z) \] \, d t \ + \ \s_1 (\rho,\vartheta,z) \, d w_2 \label{eq:ex_cylinder} \\
d z &=& \gamma (r,\vartheta,z) \, d t \ + \ \s_2 (\rho,\vartheta,z) \, d w_3 \ , \nonumber  \end{eqnarray}
with $\{ a,b,c,\kappa, \om \}$ nonzero real constants and $\{\ga , \s_1 , \s_2 \}$ smooth functions. By construction, the cylinder $M_0$ of equation $\rho = 1$ is an invariant submanifold (the function $J_1 = \rho$ is a conditional invariant), and on it the system reduces to
\begin{eqnarray*}
d \vartheta &=& \[ \omega \, + \, c \, \cos (z) \] \, d t \ + \ \s_1 (1,\vartheta,z) \, d w_2 \\
d z &=& \gamma (1,\vartheta,z) \, d t \ + \ \s_2 (1,\vartheta,z) \, d w_3 \ , \end{eqnarray*}

Suppose now that in general $\{\ga,\s_1,\s_2\}$ depend effectively on $\vartheta$, but that
$$ \( \pa_\vartheta \ga (\rho,\vartheta,z) \)_{\rho=1} \ = \ \( \pa_\vartheta \s_1 (\rho,\vartheta,z) \)_{\rho=1}  \ = \ \( \pa_\vartheta \s_2 (\rho,\vartheta,z) \)_{\rho=1} \ = \ 0 \ . $$
Then the restricted system admits rotations in $\vartheta$ as symmetries, and hence the system can be further reduced to an equation in $z$. Albeit we are in general not able to solve this (one-dimensional) equation, we know that given a solution we can reconstruct a solution of the two-dimensional restricted system, and this is a special solution to the full three-dimensional problem.

Similarly it is easy to produce examples in which not only the cylinder $\rho = 1$ is invariant, but within this we have the circle $S^1_0 = \{ \rho = 1 \ , \ z=0 \}$ which is also invariant. In this case the equation is fully integrated for initial data on $S^1_0$, partially integrated for initial data on the cylinder $\rho=1$, and in general nothing can be said for general initial data (but Eq. \eqref{eq:ex_cylinder} show that an asymptotic analysis will be possible). \EOE

\medskip\noindent
{\bf Example \ref{sec:examples}.4.} In the previous examples, everything was rather obvious by construction, thanks to the use of polar coordinates and the fact the invariant manifold was just a circle of radius one. Things are slightly less evident if we use Cartesian coordinates, and we want to show now that one can indeed work also with a less ``ready to use'' setting.

Consider, with $r^2 = x_1^2 + x_2^2$ for short, the equation
\begin{eqnarray}
dx_1 &=&  \[ \( \a \, (1 - r^2 ) \, - \, (\s^2 / 2) \) \, x_1 \ - \ \( \b \, (1 - r^2) \, + \, \om \, r^2 \) \, x_2 \] \ d t \nonumber \\ & & \ + \ \s \, (1 - r^2 ) \, x_1 \, d w_1 \ - \ \s \, x_2 \, dw_2 \nonumber \\
dx_2 &=& \[ \( \b ( 1 - r^2 ) \, + \, \om \, r^2 \) \, x_1 \ + \ \( \a \, (1 - r^2 ) \, - \, (\s^2 / 2) \) \, x_2 \] \ d t \nonumber \\ & & \ + \ \s \, (1 - r^2) \, x_2 \, dw_1 \ + \ \s \, x_1 \, dw_2 \ . \label{eq:excart65} \end{eqnarray}

We consider now the function $J = r^2$, and look at its variation on the dynamics \eqref{eq:excart65}. This yields
\begin{eqnarray}
dJ &=& \frac{\pa J}{\pa t} \, dt \ + \ \frac{\pa J}{\pa x_1} \, d x_1 \ + \ \frac{\pa J}{\pa x_2} \, d x_2 \ + \ \frac12 \, \Delta (J) \, d t \nonumber \\
&=& \[ \( 2 \, a \ + \  (1 - J) \, \s^2 \) \, J \, (1 - J) \] \ d t \ + \ 2 \, \s \, J \, (1 - J) \, d w_1 \ . \end{eqnarray}
Thus we obtain that $J$ is \emph{not} an invariant; but it is a \emph{conditional invariant}: on the level sets $J=0$ (corresponding to the origin in the  Cartesian plane) and $J=1$ (the unit circle in the Cartesian plane) we have $d J=0$, and hence these are invariant manifolds for the dynamics \eqref{eq:excart65}.

Actually, passing to polar coordinates $(\rho, \theta)$ (on $\rho = 0$ the change of coordinates is singular, and $\theta$ ill-defined) the Eq \eqref{eq:excart65} reads simply
\begin{eqnarray}
d \rho &=& \a \, \rho \, (1 - \rho^2 ) \, dt \ + \ \s \, \rho \, (1 - \rho^2) \, d w_1 \nonumber \\
d \theta &=& \[ \b (1 - \rho^2 ) \, + \, \om \, \rho^2 \] \, dt \ + \ \s \, dw_2 \ . \label{eq:excart65polar} \end{eqnarray}
It is thus evident that on $J=\rho^2 = 1$ the dynamics reduces to
\begin{eqnarray*}
d \rho &=& 0 \\
d \theta &=& \om \, dt \ + \ \s \, dw_2 \ . \end{eqnarray*}

It should be noted that, as immediately apparent from \eqref{eq:excart65polar}, in this case rotations in the $(x_1,x_2)$ plane are a symmetry of the full equation, and not only of the reduced one. We thus see again that one can have a conditional invariant associated to a full symmetry. \EOE

\medskip\noindent
{\bf Example \ref{sec:examples}.5.} We generalize the previous Example; we use Cartesian coordinates $(x,y)$ and write again $r = \sqrt{x^2 + y^2}$. Consider the equations
\begin{eqnarray}
d x_1 &=& \[ \( (1 - r )  x_1  f_1  -  x_2  f_2 \)  - \frac{(1 - r)^2}{2} \( (s_{21}^2 + s_{22}^2 )  x_1 + 2 (s_{11} s_{21} + s_{12} s_{22} )  x_2 \) \] \ d t \nonumber \\
& & \ + \ (1-r)  (s_{11}  x_1  -  s_{21}  x_2 ) \, d w_1 \ + \ (1 - r) (s_{12}  x_1  -  s_{22}  x_2 ) \, d w_2 \nonumber \\
d y &=& \[ \( x_1  f_2 + (1-r)  x_2  f_1 \) + \frac{(1-r)^2}{2} \( 2 (s_{11} s_{21} + s_{12} s_{22} ) x_1 - (s_{21}^2 + s_{22}^2 )  x_2 \) \] \ d t \nonumber \\
& & \ + \ (1-r)  \( s_{21} x_1 + s_{11} x_2 \) \ d w_1 \ + \
(1-r) \( s_{22} x_1 + s_{12} x_2 \) \ d w_2 \ . \label{eq:cartgen} \end{eqnarray}
Here $f_i = f_i (x,y)$ and $s_{ij} = s_{ij} (x,y)$ are arbitrary smooth functions, and of course the system is written in this rather involved way so that it produces a simple result.

We consider again the function $J= r^2$; it results that on solutions to \eqref{eq:cartgen} we have
\begin{eqnarray*}
d J &=& \[ (1 - r) \, J \, \( 2 f_1 (x,y) + s_{11}^2 + s_{12}^2 - r (s_{11}^2 + s_{12}^2) \) \] \, dt \\
& & \ + \ 2 \, (1 - r) \, J \, s_{11} \ d w_1 \ + \ 2 \, (1 - r) \, J \, s_{12} \, dw_2 \ . \end{eqnarray*}
It is thus immediately apparent that $J$ is in general \emph{not} an invariant for \eqref{eq:cartgen}, but that it is always a conditional invariant on the circle $r=1$. Thus the latter is an invariant manifold for \emph{all} the equations of the form \eqref{eq:cartgen}.

In fact, passing to polar coordinates $(r,\theta)$, and writing $$ F_i (r , \theta ) = f_i [x_1 (r,\theta) , x_2 (r,\theta)] \ , \ \ \ S_{ij} (r , \theta ) = s_{ij} [x_1 (r,\theta) , x_2 (r,\theta)] \ , $$ the Eq \eqref{eq:cartgen} reads simply
\begin{eqnarray}
d r &=& r \ (1 - r) \ \[ F_1 (r,\theta) \ d t \ + \ S_{11} (r,\theta) \ d w_1 \ + \ S_{12} (r,\theta) \ d w_2 \] \nonumber \\
d \theta &=& F_2 (r,\theta) \ dt \ + \ (1 - r) \ \[ S_{21} (r,\theta) \ d w_1 \ + \ S_{22} (r,\theta) \ d w_2 \] \ . \label{eq:cartgenpol} \end{eqnarray}
Now, unless all the functions $F_i$ and $S_{ij}$ are actually independent of $\theta$, rotations are \emph{not} a symmetry of the full Eq \eqref{eq:cartgenpol}, and hence of the original Eq \eqref{eq:cartgen}.

On the other hand, the equation reduced to the invariant manifold $r=1$ reads just
\beq dr \ = \ 0 \ , \ \ \ d \theta \ = \ F_2 (1, \theta) \ d t \eeq
(note this is a deterministic equation), and \emph{if} $F_2 (1,\theta)$ is independent of $\theta$, it has a rotational symmetry. On the other hand, if $F_2 (1,\theta)$ does actually depend on $\theta$, we do not have a rotational conditional symmetry, despite the presence of the conditional invariant $J$ on $r=1$. \EOE

\subsection{Asymptotic symmetries}

The reader has surely noted that, albeit we only discussed conditional symmetries and/or invariants in our previous Examples, these are actually tailored so to have asymptotic symmetries and/or invariants. We will look again at them under this point of view, but we will first consider an even simpler one dimensional Example.

\medskip\noindent
{\bf Example \ref{sec:examples}.6.} Consider the
one-dimensional equation \beq d x \ = \ - \, \left( x \
+ \ x^3 \right) \ d t \ + \ \sigma \, x \, d w \ ; \eeq in this case
the invariant manifold is just the single point $x=0$; this is of course of little interest in
itself.

This equation does not have continuous symmetries (an obvious reflection symmetry $(x,t,w) \to (- x,t,w)$ is present), but as the dynamics is attracted
towards $x=0$ we know that asymptotically the system dynamics will
be described by its linearization \beq d x \ = \ - \, x \ d t \ +
\ \sigma \, x \ d w \ . \eeq This has a scaling symmetry, \beq X \
= \ x \ \frac{\partial}{\partial x} \ := \ \varphi (x) \
\frac{\partial}{\partial x}\ ; \eeq thus we can just use the
prescription discussed in Sect. \ref{sec:symmsto} and pass to the
new variable
$$ y \ = \ \int \frac{1}{\varphi (x)} \ d x \ = \ \log (x) \ . $$

Note here that we should actually write $y = \log (|x|)$ and
distinguish the cases $x>0$ and $x<0$. This is just fine as $x=0$
is a barrier for this dynamics. We will then use $x > 0$ (so that in particular $\pa y / \pa x = 1/x$), and $x = \exp[y]$. A similar discussion would apply for $x < 0$.

If we apply this transformation to the full equation, we get
\begin{eqnarray*} d y &=&
\frac{1}{x} \ d x \ + \ \frac12 \, \Delta (y) \, d t  \\ &=& \frac{1}{x} \ \left[ - (x + x^3) \ dt \
+ \ \sigma x \, d w \right] \ - \ \frac12 \, \s^2 \, dt \\
&=& - \, (1 + x^2 + \s^2/2) \, dt \ + \ \s \, d w \\
&=& - \, \[ \(1 + \s^2/2\) \, + \, e^{2 y} \] \, dt \ + \ \s \, d w \ . \end{eqnarray*} Now we should observe that we are interested in the region $x \approx 0$, i.e., $y \approx - \infty$; in this limit the term $e^{2 y}$ can be discarded. Actually, the $y$ dynamics will drive the system towards this limit.  Thus in this asymptotic regime the dynamics is just governed by
$$ d y \ = \ - \( 1 + \s^2/2 \) \, dt \ + \ \s \, dw \ ; $$
it is clear that this equation is readily integrated to
$$ y(t) \ = \ y(t_0) \ + \ \( 1 + \s^2/2 \) \, (t-t_0) \ + \ [ w(t) - w(t_0)] \ , $$
and also that it admits $\pa_y$ as a symmetry.
\EOE

\medskip\noindent
{\bf Example \ref{sec:examples}.7.} Consider again the system of Example \ref{sec:examples}.2, i.e., \eqref{eq:ex_6_1}, which we rewrite here for ease of reference:
\begin{eqnarray}
d \rho &=&  a \, (1 - \rho^2) \, \rho \, d t \ + \ \s \, (1 - \rho^2) \, d w_1 \nonumber \\
d \vartheta &=& \[ b \, (1 - \rho^2) \ + \ \om \, \rho^2  \] \ d t
\ + \ \s \, d w_2 \ . \label{eq:ex96} \end{eqnarray}
The equation for $\rho$ is autonomous, and we readily note that $\rho = 1$ is an invariant manifold, and actually for $a > 0$ a weakly attractive one.

It appears from \eqref{eq:ex96}  that -- at least for $|\s| \ll 1$, i.e. when the drift overcomes the diffusion term -- the dynamics converges to the circle $\rho = 1$. Note this will be a weakly attracting manifold. We will thus linearize \eqref{eq:ex96} around $\rho = 1$; we do this by writing
$$ \rho = 1 + \eta $$ and taking only first order terms in $\eta$. This yields
\begin{eqnarray}
d \eta &=& - 2 \, a \, \eta \, dt \ - \ 2 \, \s \, \eta \, dw_1  \nonumber \\
d \vartheta &=& \[ \om \ + \ 2 \, \eta (\om \, - \, b ) \] \, dt \ + \ \s \, dw_2 \ \label{eq:ex96red}. \end{eqnarray}

The dynamics is attracted to $\eta = 0$ (corresponding to $\rho =1$) and on this it is described by $d \vartheta = \om dt + \s dw_2$, while in the region around the attracting manifold it is described by \eqref{eq:ex96red}.

The original Eq \eqref{eq:ex96} is rotationally invariant; this symmetry is also present in the reduced Eq \eqref{eq:ex96red}.

Note also that the first equation in \eqref{eq:ex96red} admits $X = \eta \pa_\eta$ as a (scaling) symmetry; thus it can be integrated (passing to the variable $y = \log (\rho) = \log (1 + \eta)$), which yields $\eta = \eta (t)$ as an explicit function, depending on the realization of the Wiener process $w_1 (t)$. With this, the second equation in \eqref{eq:ex96red} is also readily integrated.

It may be interesting to see what happens if we proceed in a slightly different way. We can perform the change of variables $y = \log (\rho)$ directly on the original equation \eqref{eq:ex96}; this yields in mixed notation (obviously we have a singularity at $\rho = 0$) for the first equation
$$
dy \ = \ \[ \( a \, - \, \frac{\s^2}{2 \, \rho^2} \, + \, \frac{\s^2}{2}  \) (1 - \rho^2) \] \ dt \ + \ \s \, \frac{1 - \rho^2}{\rho} \ d w_1 $$
Passing now to write $\rho = 1 + \eta$ as above, the full equation reads
\begin{eqnarray*}
d y &=& - \ \[ \eta \, (2 + \eta ) \, \( a \, - \, \frac{\eta (2 + \eta) \s^2}{2 \, (1 + \eta)^2} \) \] \ dt \ - \ \s \, \eta \, \frac{2 + \eta}{1 + \eta} \ d w_1 \ ;  \end{eqnarray*}
expanding in series at first order in $\eta$, we get $y = \log (1 + \eta) \simeq \eta$, hence $dy \simeq d \eta$,  and
$$ d \eta \ = \  - \, 2 \, a \, \eta \, dt \ - \ 2 \, \s \, \, \eta \, d w_1  \ , $$
where $|\s | \ll 1$ has been taken into account, exactly as before. \EOE

\medskip\noindent
{\bf Example \ref{sec:examples}.8.} Consider the equation
\begin{eqnarray}
d x_1 &=&  \[ (1 - r) \, \( a \, x_1 \ - \ b \, x_2 \) \ - \ \om  \ x_2  \ - \ (\a^2 + \b^2 ) (x_1/2) \] \ d t \nonumber \\
& & \ - \ \a \, x_2 \, d w_1 \ - \ \b \, x_2 \, d w_2 \nonumber \\
d x_2 &=& \[ (1-r) \, \( b \, x_1 \ + \ a \, x_2 \) \ + \ \om \, x_1 \ - \ (\a^2 + \b^2 ) \, (x_2 / 2) \] \ d t \nonumber \\
& & \ + \ \a \, x_1 \, d w_1 \ + \ \b \, x_1 \, d w_2 \ . \label{eq:exgenasy} \end{eqnarray}
Here $a,b$ are real constants, $\a,\b,\om$ are arbitrary smooth functions of $x_1,x_2$. We assume $a > 0$ for definiteness.

In this case the evolution of $J = r^2$ is given by
$$ d J \ = \ 2 \, a \ r^2 \, (1 - r) \ d t \ ; $$
note this is a deterministic equation, so that any initial datum with $r \not= 0$ is attracted by the circle $r=1$. In this case, we have an invariant and \emph{strongly} attractive submanifold. Correspondingly, $J$ is a \emph{strongly asymptotic invariant}.

Passing to polar coordinates, we have
\begin{eqnarray*}
d r &=& a \ r \ (1 - r) \ dt \\
d \theta &=& \[ b \, (1-r) \ + \ \om \] \ d t \ + \ \a \ d w_1 \ + \ \b \ d w_2 \ . \end{eqnarray*}
Needless to say, the functions $\a,\b,\om$ should now be thought as functions of $(r,\theta)$; we will denote by $\^\a, \^\b, \^\om$ the restrictions of these to the circle $r=1$. On the invariant strongly attractive submanifold $r = 1$ the evolution is described by
\beq d \theta \ = \ \^\om (\theta) \ d t \ + \ \^\a (\theta ) \ dw_1 \ + \ \^\b (\theta) \ d w_2 \ . \eeq
It is rather clear that choosing suitably the functions $\a, \b,\om$ we can have different situations:
\begin{itemize}
\item If these do not depend on $\theta$, the full Eq \eqref{eq:exgenasy} is rotationally invariant, i.e., $X = - x_2 \pa_1 + x_1 \pa_2$ is a full symmetry of \eqref{eq:exgenasy};
\item If these do depend on $\theta$, but their restrictions $\^\a, \^\b , \^\om$ do not depend on $\theta$, then  \eqref{eq:exgenasy} does not admit $X$ as a full symmetry, but admits it as a conditional symmetry on the invariant (and attractive) manifold $M_0$ identified by $r=1$;
\item If (at least one of) the $\^\a, \^\b , \^\om$ depend on $\theta$, then \eqref{eq:exgenasy} does not even admit $X$ as a conditional symmetry.
    \end{itemize}

It is maybe worth stressing again that in all these cases $J$ is a strongly  \emph{asymptotic invariant}. \EOE

\section{Discussion and conclusions}

In many physically interesting situations, one observes that equations not enjoying a given symmetry can have special solutions which are symmetric. Even more frequently, the asymptotic behavior of the system under study is characterized in terms of symmetry, independently of the symmetry of initial conditions; typically this involves a simple, i.e., a rotational or a scaling symmetry. A theory of conditional and asymptotic symmetries for \emph{deterministic} equations has been developed in the literature.

In the present paper, we have extended this theory to the case of \emph{stochastic} differential equations, considering also the interplay between (asymptotic) symmetry and (asymptotic) integrability for stochastic equations. A special role in this frame is played by (full or conditional or asymptotic) \emph{invariants} for stochastic equations. We have also shown by concrete (simple) examples how the theory developed here can be applied in practice.

Our examples show that the interrelation between the presence of (conditional and/or asymptotic) symmetry and invariants is not unique: the presence of conditional of asymptotic configurational invariants is needed to have a conditional or asymptotic symmetries, but the converse is not true.

Our discussion shows that the pre-existing theory for deterministic equations can be fully extended to stochastic ones, and the examples show that it can be applied with a well definite algorithm and standard computations; these were specially simple in the considered examples -- which were built precisely in order to show the conceptual issues without blurring them by computational details -- but our discussion shows that even in more involved cases (apart from computational complications) the procedure is rather straightforward.

Needless to say, its success depends on the equations under study possessing the required (full, conditional or asymptotic) symmetry properties. Symmetry is a non-generic property; but conditional or even more asymptotic symmetry is more common than full symmetry, so we are confident that our method can be widely applicable, as it already happens in the context of deterministic equations.

Our approach was focusing -- again, as in the deterministic case -- on \emph{continuous} (Lie) symmetries. At difference with the deterministic case, we have dealt only with ordinary stochastic differential equations, and not discussed partial stochastic differential equations. The reason for this is that a symmetry theory for these equations is not as well developed as for ordinary ones; we expect that once this (substantial) obstacle is removed, the theory can be extended from full to conditional and asymptotic symmetries as in this work. In view of the relevance of conditional and asymptotic symmetries for the study of deterministic PDEs, we expect this extension to stochastic PDEs to be specially relevant for applications, and we hope it may soon be obtained.

Finally, we would like to note that in a way the approach pursued in the literature dealing with conditional or asymptotic symmetries is inverse to that of the well known \emph{symmetry breaking} approach. In that case, attention is focused on the fact that equations possessing a given symmetry $G$ can have solutions which are invariant only under a subgroup $G_0 \ss G$ (possibly with $G_0 = \{ I \}$); here instead one is focusing on the fact that an equation possessing a symmetry $G_0$ (possibly with $G_0 = \{ I \}$) can have solutions with symmetry $G \supset G_0$. Similarly, for an equation with symmetry $G_0$ it may happen that the asymptotic dynamics is described by an effective equation with symmetry $G \supset G_0$, and this is turn may have solutions with symmetry $G$ or even $\wt{G} \supset G \supset G_0$.


\addcontentsline{toc}{section}{Acknowledgements}

\section*{Acknowledgements}

The work of GG is also supported by SMRI, which he was visiting while the first draft of this paper was prepared and again at the time of final writing.
The work of FS is supported by ERC SG CONSTAMIS grant. GG and
FS are also members of GNFM-INdAM. We thank these Institutions
for their support.

\addcontentsline{toc}{section}{References}


\end{document}